\documentclass{vgtc}         

\ifpdf
  \pdfoutput=1\relax                   
  \usepackage{graphicx}                
  \DeclareGraphicsExtensions{.pdf,.png,.jpg,.jpeg} 
\else
  \usepackage{graphicx}                
  \DeclareGraphicsExtensions{.eps}     
\fi%

\graphicspath{{jpg/}{./}}

\usepackage{microtype}                 
\PassOptionsToPackage{warn}{textcomp}  
\usepackage{textcomp}                  
\usepackage{mathptmx}                  
\usepackage{times}                     
\usepackage{cite}                      
\usepackage{tabu}                      
\usepackage{booktabs}                  
\usepackage{color}
\usepackage{amssymb}
\usepackage{stmaryrd}
\usepackage{amsmath}
\usepackage{gensymb}
\usepackage{arydshln}
\usepackage{accents}
\usepackage{wrapfig}

\DeclareFontFamily{U}{mathx}{\hyphenchar\font45}
\DeclareFontShape{U}{mathx}{m}{n}{
      <5> <6> <7> <8> <9> <10>
      <10.95> <12> <14.4> <17.28> <20.74> <24.88>
      mathx10
      }{}
\DeclareSymbolFont{mathx}{U}{mathx}{m}{n}
\DeclareFontSubstitution{U}{mathx}{m}{n}
\DeclareMathAccent{\widecheck}{0}{mathx}{"71}

\onlineid{1016}

\vgtccategory{Research}

\vgtcinsertpkg

\vgtcpapertype{Application}

\newcommand{\julien}[1]{\textcolor{black}{#1}}
\newcommand{\maxime}[1]{\textcolor{black}{#1}}
\newcommand{\maximed}[1]{\textcolor{black}{#1}}
\newcommand{\maximet}[1]{\textcolor{black}{#1}}
\newcommand{\Julien}[1]{\textcolor{black}{#1}}

\newcommand{\julienWed}[1]{\textcolor{black}{#1}}
\newcommand{\maximej}[1]{\textcolor{black}{#1}}

\newcommand{\julienLDAV}[1]{\textcolor{black}{#1}}

\newcommand{\maximeLDAV}[1]{\textcolor{black}{#1}}

\title{
Ranking Viscous Finger Simulations to an Acquired Ground Truth \\
with Topology-aware Matchings}

\author{Maxime Soler, Martin Petitfrere, Gilles Darche, 
Melanie Plainchault, Bruno Conche and Julien Tierny
}

\nocopyrightspace

\newcommand{\eqrref}[1]{Eq.~\ref{#1}}
\newcommand{\figref}[1]{Fig.~\ref{#1}}
\newcommand{\secref}[1]{Sec.~\ref{#1}}
\newcommand{\tabref}[1]{Tab.~\ref{#1}}

\newcommand{\domain}{\mathcal{M}}
\newcommand{\range}{\mathbb{R}}

\newcommand{\sub}[1]{f^{-1}_{-\infty}(#1)}
\newcommand{\sur}[1]{f^{-1}_{+\infty}(#1)}

\newcommand{\persistentDiagram}[1]{\mathcal{D}(#1)}
\newcommand{\Index}{\mathcal{I}}

\newcommand{\wasserstein}[1]{W_{#1}}
\newcommand{\liftedWasserstein}[1]{\widehat{W}_{#1}}

\newcommand{\vcaption}[1]{\vspace{-3ex}\caption{#1}\vspace{-1ex}}

\newlength{\dhatheight}

\newcommand{\otherOverlap}[1]{\widecheck{O}_{#1}}
\newcommand{\otherOverlapIndex}[1]{\widecheck{O}_{#1}}

\newcommand{\otherLiftedWasserstein}[1]{\widecheck{W}_{#1}}
\newcommand{\otherLiftedWassersteinIndex}[1]{\widecheck{W}_{#1}}

\newcommand{\removed}[1]{}

\abstract{
\maxime{
This
\julienLDAV{application}
paper presents a \Julien{novel} framework \Julien{based on topological 
data analysis} for the automatic evaluation and ranking
of viscous finger simulation runs in an ensemble with respect to a reference
acquisition.
Individual fingers in a given time-step are associated with critical point pairs
in the distance field to the injection point, forming persistence diagrams.
Different metrics, based on \Julien{optimal} transport, for comparing 
time-varying persistence diagrams
in this specific applicative case are introduced.
We evaluate the \Julien{relevance} of the rankings obtained with these 
metrics,
both qualitatively thanks to a lightweight web visual interface, 
and quantitatively by studying the deviation from a reference ranking suggested
by experts. 
\Julien{Extensive experiments show the quantitative superiority
of our approach compared to traditional alternatives.}
Our web interface allows experts to conveniently 
explore the produced rankings.
We show a complete viscous fingering case study demonstrating the utility
of our approach in the context of porous media fluid flow\Julien{, where our 
framework can be used to automatically discard physically-irrelevant 
simulation runs from the ensemble and rank the most plausible ones}.
We \Julien{document} an in-situ implementation to lighten
I/O and performance constraints arising in the context of parametric studies.
}
} 

\teaser{
  \centering
  \includegraphics[width=\linewidth]{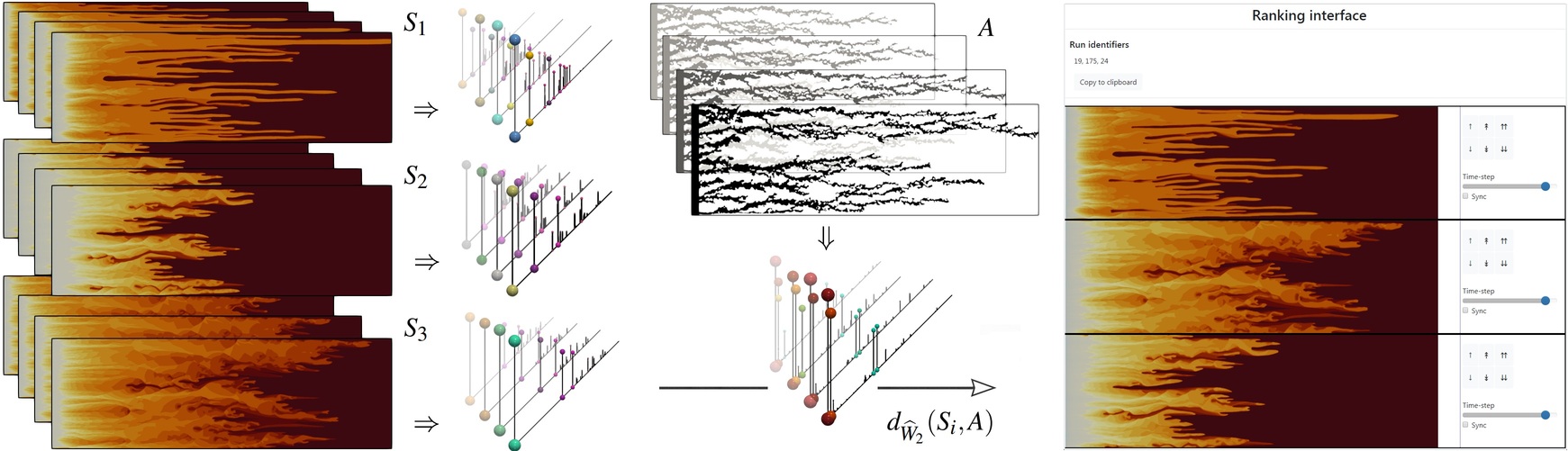}
  \vcaption{\maximet{Overview of the ranking framework. An ensemble of 
   viscous fingering simulation runs $S_1$, $S_2$, $S_3$ is launched, and the 
   persistence diagram of newly available time-steps can be computed 
   \emph{in-situ} (left). Only persistence diagrams for
   which there is a matching ground truth image (center, top) are computed.
   Diagrams of every simulation are compared with the diagrams of the
   ground truth (center, bottom) at matching time-steps. 
   This comparison, based on a metric $\liftedWasserstein{2}$  
   combining the notions of persistence and geometry, 
   outputs a distance measurement, which can be integrated over time to form
   the metric $d_{\liftedWasserstein{2}}$. This produces a
   final ranking (right) which characterizes the quality of simulations,
   allowing experts to select and explore best performing runs automatically. } }
	\label{fig:teaser}
}

\begin{document}


\firstsection{Introduction}

\maketitle


The chaotic nature of fluid flows makes it difficult to account for
the propagation of initial uncertainties in numerical models,
or uncertainties in model parameters.
To predict 
uncertain phenomena, thanks to the increase
in computing power in recent years,
Monte Carlo methods have been broadly used, for instance 
in climate modeling, forecasts, 
statistical physics, chemistry and 
astrophysics. The idea is to compute a large number of
simulations, called an \emph{ensemble},
while \maxime{densely} 
sampling the space of input parameters.
A \emph{post-mortem} comparison
\maxime{(i.e. performed \emph{after} simulations have been completed)}
\maxime{to} 
experimentally acquired data 
can then determine \maxime{which simulations produced the most realistic outcomes}
and how 
input parameters affect their variability.

Specifically, in reservoir engineering,
an area of petroleum engineering
concerned with fluid flow through porous media,
it is important to quantitatively predict well productions,
i.e. the quantity of oil that can be extracted. 
\maximeLDAV{For example, injector wells are commonly drilled 
to inject water and flush the oil in place, which is extracted by producer wells; 
simulation can help well placement optimization and production forecast.} 
Numerical models 
are subject to parameter uncertainties, and can be tuned
by launching randomly sampled ensemble simulations.
Usually, reference production \maxime{rates and well pressures are} 
history-matched with 
the simulated ensembles, which ideally would allow domain experts
to restrict the space of input parameters.
\maximej{This history match procedure is usually applied at the field scale 
(oil and gas reservoirs), but also at the core scale (a few decimeters)
when lab engineers want to match the behavior of experimental corefloods.} 

In practice, 
notably in the domain of Darcy-type simulations
at the core scale,
production \maxime{and pressure} data is not 
sufficient to infer model parameters.
\maxime{Further \maximed{measuring} 
tools have been recently integrated
in lab experiments in order to constrain the parameter space, 
by monitoring the \emph{saturation} scalar fields through X-rays,
so as to obtain information on phase velocities and residual saturations.
Here the saturation measures the volume fraction of a \maximed{given phase} 
in the geometrical domain.}
Observing these scalar fields seems relevant when the fluid
behaves in a particularly \maximed{chaotic} 
way, so that 
simulations which are not physically adequate could be detected.
The case of the viscous fingering phenomenon\maxime{, an 
instability which occurs at the interface between two fluids
of distinct viscosity in porous media,}
is of \maxime{particular} 
interest. 

\maxime{In this context, 
\maximed{for all simulations quantitatively reproducing production and pressure data,
experts have to visually inspect each member of the ensemble to further
discard non-physical simulations. This process is currently performed
manually and can be time consuming.} 
Moreover, the viscous fingering
process involves a notoriously chaotic and unstable geometry.
In particular, two different viscous fingering simulations can both be
realistic from an expert's point of view (and yield valid physical
properties for reservoir exploitation)
even though saturation would admit fingers with a drastically
different shape and distribution in space.
This high geometrical variability makes it particularly
challenging to derive a meaningful distance metric to compare
saturation scalar fields between a simulation and a ground truth.}

For studying scalar fields, topological data analysis \maxime{(TDA)} has been 
used in recent years as a robust
and reliable setting, allowing 
\julienLDAV{one}
to hierarchically define features
of interest in the data \cite{edelsbrunner09}. Its applicability 
to time-varying data \cite{bajaj06, bremer10},
ensembles \cite{favelier2019persistence}
and comparisons \cite{soler2018lifted}
make\Julien{s} it a reliable candidate for assessing the likeliness of 
simulations in an ensemble given a ground truth.
\maxime{
Although several approaches have explored the 
promising potential of TDA for extracting and
characterizing the features of interest in viscous
fingering simulations \cite{favelier2016visualizing, lukasczyk2017viscous},
no approach has been proposed to estimate the similarity
between two time-varying viscous fingerings based on  
topological representations.  
}

\maxime{
In this \julienLDAV{application} paper, we address the aforementioned issues by 
proposing
a novel framework, based on topological data analysis, for quantitatively ranking
simulations from an ensemble with respect to a ground truth in
a viscous fingering case study.
This framework allows experts to easily separate
the most realistic simulations from
the most unrealistic ones.
It is based on a new approach for comparing temporal sequences of 
\emph{persistence diagrams}, specifically adapted to the problem
of viscous fingering. Extensive experiments quantitatively show the superiority
of our approach compared to traditional alternatives. 
The framework also includes an interactive \Julien{visual} system for exploring 
the output rankings.
Finally, we \Julien{report} a complete case study for which the presented 
approach has been
applied \emph{in-situ} (i.e. during the simulation).
}

\subsection{Related work}

\maxime{
Since the work presented in this \Julien{paper} involves multiple
 domains (viscous finger simulations, topological data analysis, 
time-varying data), this section briefly 
\Julien{presents their respective related work.}
}

\noindent
\textbf{Viscous fingering} is a well-known instability
encountered in soils and porous 
media \cite{saffman1958penetration}, arising from the 
\maxime{unfavorable mobility ratio between an injected fluid and 
the fluid in place, for instance when injecting water in a highly
viscous oil.} 
These phenomena have been studied in the context
of petroleum engineering at multiple scales 
\cite{skauge2014polymer, gao2011advances}. Other factors
than the viscosity ratio are at play, such as properties inherent
to the medium in which the fingering takes place
\cite{homsy1987viscous, trojer2015stabilizing, skauge2011experimental, 
tsuji2016characterization}.
\maxime{In practice, performing waterflood in highly viscous oil can
lead to physical instabilities resulting in fingering patterns, with
water flowing in preferential paths and bypassing large quantities of oils.}
\maxime{To prevent this phenomenon, polymer can be injected in order to increase
the water viscosity\maximed{, therefore making the injection front more stable,
and leading to increased macroscopic oil recovery}} \cite{de2018numerical}.
There are multiple numerical models
that can describe the evolution of fluids in the context of water 
floods;
some have been qualitatively compared to acquisitions
\cite{riaz2007forced, sharma2012experiments, de2018numerical},
but the literature lacks robust way\maxime{s} to 
\maxime{quantitatively compute their difference.}

\noindent
\textbf{In-situ: }
as current trends in super-computing indicate an
increase of the computing power that evolves faster than
memory, IO and network bandwidth, new paradigms for scientific 
simulation are needed.
\maxime{
The simulation of flow in porous media, of key importance
for studying viscous fingers, is particularly affected by
data movement problematics, as models keep increasing in size, 
and high-resolution time sampling is required for observing realistic simulations.}
Over recent years, solutions for limiting data movement
were developed in this perspective, such as
\emph{in-situ} \maxime{\cite{yu10, Rivi2011InsituVS, Rasquin11, 
o2016cinema, ayachit2016performance}} and \emph{in-transit} 
\maxime{\cite{bennett12, Moreland2011}} \julien{models},
with a clear ambition to
reach toward exascale computing \cite{son14, chris13} in the 
forthcoming years.
\maxime{\Julien{However, to}
the best of our knowledge, no data analysis 
method 
has yet been proposed 
\Julien{for the in-situ analysis of viscous fingers.}}

\noindent
\textbf{Topological data analysis (TDA) techniques}
\cite{edelsbrunner09,
pascucci_topoInVis10, heine16, Defl15} 
have been used 
over the course of recent years because of their 
ability to hierarchically identify features in scalar data 
in a generic, robust
\cite{edelsbrunner02, cohen-steiner05} and efficient
manner.
They have been applied in various scientific domains, such as
computational fluid dynamics \cite{kasten_tvcg11, favelier2016visualizing},
turbulent combustion \cite{bremer_tvcg11},
material sciences \cite{gyulassy_vis15},
\julien{biological} imaging \cite{carr04, bock18},
chemistry \cite{harshChemistry, chemistry_vis14}, 
astrophysics \cite{shivashankar2016felix, sousbie11},
ensemble clustering \cite{favelier2019persistence}, 
compression \cite{soler2018topologically} or
feature tracking \cite{soler2018lifted}.
\Julien{One of the reasons for the successful applications  of TDA is}
the possibility for experts to
easily translate high-level domain-specific notions
in terms of topological data structures,
which are abstractions related to geometrical
aspects or discrepancies in the data.
Among such abstractions are
persistence diagrams \cite{edelsbrunner08b, edelsbrunner02}, 
contour trees \cite{carr2003computing}, Reeb \julien{g}raphs
\cite{pascucci07, biasotti08, tierny_vis09}, Morse-Smale complexes 
\cite{gyulassy_vis08}.
For instance, in astrophysics the cosmic web can be extracted by querying the
most persistent 1-separatrices of the Morse-Smale complex
connected to maxima of matter density \cite{sousbie11}.
Similar 
\julien{TDA applications}
can be found in the above examples.
Topological data analysis techniques have \julien{also} been applied 
\emph{in-situ}
\cite{landge14}, which demonstrates their interest and \julien{relevance}
in the context of large-scale simulations.
\maxime{
It has also been used to study the viscous fingering phenomenon,
for instance in ensembles of \julien{particle} simulations, 
to determine how the resolution
affects fingers \cite{favelier2016visualizing, lukasczyk2017viscous} 
or to provide frameworks for their visual exploration 
and interpretation \cite{luciani2019details}; but never,
to our knowledge,
for the purpose of comparing simulations, \julien{in particular} with a 
reference.}

\noindent
\textbf{\maxime{Feature-oriented distances:}} 
for comparing simple discrete scalar fields such as images, 
intuitive approaches are \maxime{point-wise} 
geometric distances such as the 
Euclidean and chord distances, or distances with a statistic awareness
such as the Mahalanobis distance or correlation coefficients
\cite{chen2005similarity}.
In specific applicative domains, however,
the experts' knowledge should be accounted for
to gain a more precise insight of what is of
actual interest in the data and which patterns or subsets
are interesting to compare.
Consequently, feature-oriented distance definitions are exposed
in the remainder of this section.
Associating geometrical loci in scalar data 
based on a high-level definition of features of interest
often relies on computing the overlap of geometrical sub-domains
\cite{Saikia17, bremer10, bremer_tvcg11, bajaj06, 
Silver95, Silver97, Silver972, Silver98}.
Such methods are used for \julien{feature} tracking in time-varying data 
\cite{silver99}.
On another note, Transportation theory 
offers an important continuous formulation of 
this problematic, with the notion of a \emph{Wasserstein} 
or \emph{Earth mover's} distance
\cite{monge81, Kantorovich42, levina01}, which has 
gained interest in recent years 
\cite{Cuturi13, solomon16, solomon17, lavenant2018dynamical}.
In the discrete setup,
when applied to topological structures such as
persistence diagrams,
transport-based matching methods suffer from 
instabilities \julien{in the geometrical domain} \cite{Cohen-Steiner06}, for 
which the underlying metric 
\julien{can}
be specifically corrected \cite{soler2018lifted} depending 
on the context.
\maxime{Though this family of approaches for computing distances
between features based on transport seem\julien{s} promising for 
the problem of comparing viscous fingers, there is, to our knowledge,
no work studying such an application.}

\subsection{Contributions}

This \julienLDAV{application} paper makes the following new contributions:
\begin{enumerate}
  \item{\textbf{Approach:} we present a \julien{novel} analysis framework 
allowing to 
   select \julien{relevant} members in a simulated ensemble given a ground 
truth.
   The system yields a ranking that allows \julienLDAV{experts} to visually 
explore the most likely
   simulations and discard the most unrealistic ones.
   }
  \item{\textbf{Metrics:} new topological metrics for comparing time-varying
   viscous fingers are introduced, based on the Wasserstein matching of 
   persistence diagrams, 
   specifically tuned for the viscous fingering phenomenon and
   integrated over time.
   }
  \item{\textbf{Case study:} a complete case study of a 
   viscous fingering simulation ensemble is \julien{documented},
   along with a proof-of-concept in-situ implementation of 
   our approach.
   }
  \item{\textbf{Evaluation:} 
   \julien{the metrics and ranking framework are qualitatively evaluated with 
feedback from domain experts. The quantitative performance of our approach is 
also analyzed and its superiority over traditional alternatives is 
demonstrated.}}
\end{enumerate}

\section{Preliminaries}

This section describes the 
context of reservoir simulation,
introduces our formal setting and the metrics \julien{which are 
extended by our \julienLDAV{work}.}
%

\subsection{Darcy-type porous media simulation}

There are multiple models for simulating flow in porous media.
\maximed{Though our viscous finger analysis framework
is not limited to a specific simulation model,
we introduce here} 
Darcy-type simulations,
for which the physics is governed by quantities averaged over control volumes.
We consider diphasic flow with oil and water.

\eqrref{eq:mass} describes mass \maxime{conservation}, 
where
$i \in \{o, w\}$ is 
the oil \maximed{and} 
water phase;
$\phi$ is the porosity of the medium; 
$\rho_i$ is the \maximed{mass} density of \maxime{phase} 
$i$; 
$S_i$ 
is the \emph{saturation} of \maxime{phase} 
$i$ (\maxime{it stands for the volume fraction of phase $i$);}
$q_i$ 
\maxime{is the well source term (injection/production) of phase $i$;}
and $\mathbf{\maxime{v}}_i$ 
is the velocity of \maxime{phase} $i$. 
\maximed{If $V_{\text{tot}}$ denotes the total volume, then 
the} 
mass of component $i$ is given by $m_i=\maximed{V_{\text{tot}}}\phi\rho_iS_i$.

\vspace{-4ex}
\begin{eqnarray}
  \frac{\partial}{\partial t} (\phi \rho_i S_i) = - \nabla \cdot \rho_i \mathbf{\maxime{v}}_i + q_i
  \label{eq:mass}
\end{eqnarray}
\vspace{-3ex}

Darcy's law is an equation that describes fluid flow in porous media, 
determined experimentally by H. Darcy in 1856 \maxime{for one phase} \cite{darcy1856fontaines},
and which can be derived from the 
\maxime{Stokes} equations \cite{whitaker1986flow}. 
It\maxime{s extension to multiphasic flow} is given in \eqrref{eq:darcy},
where $\mathbf{\maxime{v}}_i$ is the velocity of phase $i$;
$\mathbf{K}$ 
is the \maxime{absolute} permeability tensor of the
porous medium;
$\mu_i$ is the viscosity of $i$;
$\mathbf{g}$ is the acceleration of gravity;
$P_i$ is the pressure of phase $i$;
$kr_i$ 
is the relative permeability \maxime{of phase $i$}.
\maxime{In our model, $kr_i$ is} a function of water saturation.

\vspace{-3ex}
\begin{eqnarray}
  \mathbf{\maxime{v}}_i = - \mathbf{K}\frac{kr_i}{\mu_i}(\overrightarrow{\nabla}P_i - \rho_i \mathbf{g})
  \label{eq:darcy}
\end{eqnarray}
\vspace{-3ex}

Furthermore, as shown in 
\eqrref{eq:circle}, oil saturation can be simply expressed in terms of water saturation, 
and water pressure can be expressed in terms of oil pressure, with $P_c$ being
the \emph{capillary pressure}, a function of water saturation.

\vspace{-2ex}
\begin{equation}
  \begin{cases}
    S_w = 1 - S_o \\
    P_w = P_o - P_c \\
  \end{cases}
  \label{eq:circle}
\end{equation}
\vspace{-2ex}

\maxime{\julien{In this model, the} unknowns are \julien{the} saturations $S_i$ 
and pressures $P_i$. 
The system formed by \eqrref{eq:mass}, \ref{eq:darcy}, and \ref{eq:circle}
can then be solved \julien{numerically} to yield the evolution of fluid in 
porous media under Darcy's approximation.}
\julien{Moreover, models exist}
\cite{leverett1941capillary, thomeer1960introduction, brooks1964hydraulic} 
for expressing $P_c$ as a function of $S_w$, 
\maxime{which can be obtained experimentally through centrifugal fan 
experiments.}
Relative permeabilities $kr_o$ and $kr_w$, also functions of $S_w$,
are more elusive. \maximej{Numerous} models have been proposed in the literature 
in various contexts \cite{corey1954interrelation, corey1956effect, chierici1984novel, 
killough1976reservoir, carlson1981simulation, alpak1999validation, fenwick1998network}, 
and there \julien{is} a number of methods for \maximed{building them
from interpretation of lab experiments} 
\cite{oak1990three, macallister1993application, dicarlo2000three,
richardson1952laboratory, hagoort1980oil}.
Their correct definition, however, is key to a realistic description of flow in porous media,
and \maximed{can be quite difficult to obtain depending on the recovery mechanism, 
especially in processes involving 
severe viscous fingering patterns (in which case Darcy’s law can become approximate)
or when dealing with an extra fluid phase, like an injected gas phase}
\cite{baker1988three},
notably because of the limited availability of experimental measurements.
In the remainder of this work, relative permeabilities are considered as an
input parameter of simulations.

\maxime{Most of reservoir simulators are based on finite volumes discretizations
\maximed{of \eqrref{eq:mass}, \ref{eq:darcy}, \ref{eq:circle} on a gridded \maximej{2D or 3D}
model}, 
in which independent variables are constant in each grid block.}
These quantities 
must be determined at \maximed{each} 
time-step
by solving the \maximed{sets of non-linear conservation equations.} 
The results shown in the experiments section were obtained 
\maximej{in the 2D case} 
with an in-house \maxime{research reservoir} simulator
\maxime{\cite{patacchini2014four, jaure2014reservoir}} using \maxime{an}
IMPES \maxime{scheme (IMplicit Pressure, Explicit Saturation)} \cite{chen2004improved},
which separately computes saturation with an explicit
time approximation, and pressure with an implicit one.
\maxime{At every time-step, scalar data defined on control volumes is updated.
As there are multiple variables, the simulator outputs multiple fields, like phase
pressures and saturations. The pressure field is very diffusive, and in the diphasic case
the saturation is constrained by \eqrref{eq:circle}.
\julien{Thus, a}
 good indicator of the simulation state is the \julien{scalar field of water 
saturation $S_w$, which we will use as input data in the following}.}

\subsection{\julien{Persistence diagrams}}
\label{sec:diag}

\julien{This subsection describes our formal setting. It contains definitions 
adapted from \cite{ttk}. An introduction to TDA}
can be found in \cite{edelsbrunner09}.

\noindent
\textbf{Input data:} \julien{for each time step, }
the input \julien{saturation} data is considered as a piecewise
linear (PL) scalar field $f : \domain \rightarrow \mathbb{R}$ defined on a
PL $d$-manifold $\domain$ with \julien{$d = 2$} in our application.
Scalar values are given at the vertices of $\domain$ and linearly
interpolated  \julienLDAV{elsewhere.}

\noindent
\textbf{Critical points:}
if $w \in \range$ is an isovalue,
the \emph{sub-level set} of $w$,
noted $\sub{w}$, is the pre-image of the open interval $(-\infty, w)$ under $f$:
$\sub{w} = \{ q \in \mathcal{M} ~ | ~ f(q) < w \}$.
Symmetrically, the \emph{sur-level set} is $\sur{i} = \{ q \in \mathcal{M} ~ | 
~ f(q) > w \}$.
These two objects serve as segmentation tools in multiple analysis tasks 
\cite{bremer_tvcg11, carr04, bock18}.
The points $q \in \domain$ where the topology
of $\sub{f(q)-\epsilon}$ differs from that of 
$\sub{f(q)+\epsilon}$ \julien{for $\epsilon \rightarrow 0$}
are called the \emph{critical points} of $f$.
They can be classified according to their \emph{index} $\Index$: 
0 for minima, 1 for 1-saddles, $d - 1$ for $(d-1)$-saddles, $d$ for maxima.

\begin{figure}
 \centering
 \vspace{-2ex}
 \includegraphics[width=\columnwidth]{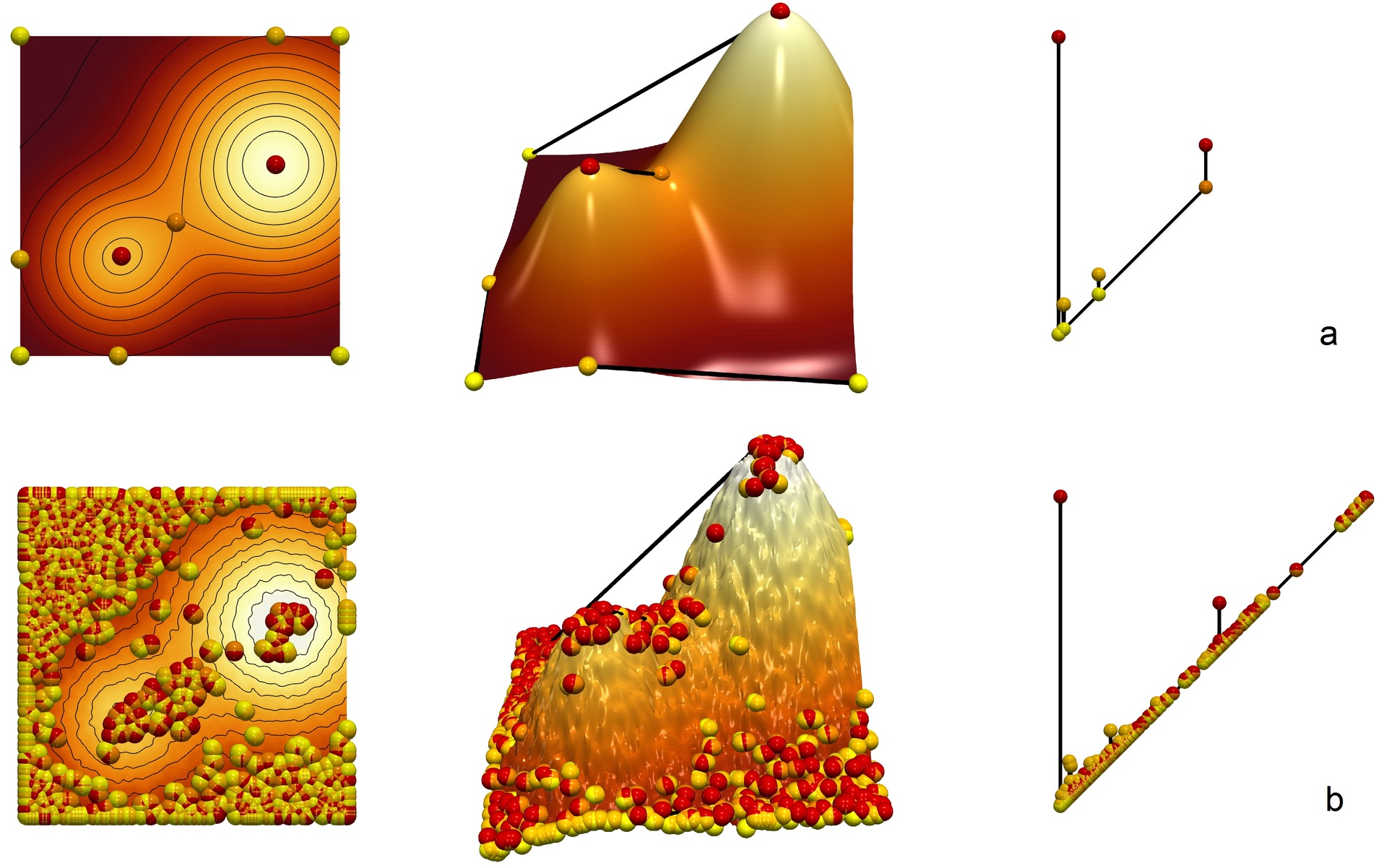}
 \vcaption{\maximet{A smooth (top row) and a noisy (bottom row) scalar field, 
   defined on a 2D domain (left), with their 3D terrain 
   representation (middle) and persistence diagrams (right).
   Critical points are represented as spheres (red: maxima, orange: saddles,
   yellow: minima). The largest pairs in the diagrams
   correspond to the two main hills.} } 
 \label{fig:fig1}
 \vspace{-2ex}
\end{figure}

\noindent
\textbf{Persistence diagrams:}
the 
\julienLDAV{set}
of critical points of $f$ can be visually 
represented by a topological abstraction called the \emph{persistence
diagram} \cite{edelsbrunner02, cohen-steiner05} (\figref{fig:fig1}).
Specifically, the topological Elder Rule \cite{edelsbrunner09} states that 
critical points can be arranged in a set of pairs, 
such that each critical point appears in only one
pair $(c_i, c_j)$ with $f(c_i) < f(c_j)$ and $\Index(c_i) = \Index(c_j) - 1$.
\julien{Such a pairing indicates that a topological feature of 
$\sub{i}$ (connected component, cycle, void, etc.) created at critical point 
$c_i$ dies at the critical point $c_j$.}
\julien{For example,}
as the value $i$ increases, if two
connected components of $\sub{i}$ meet
at a  saddle $c_j$ of $f$, the \emph{youngest} of the two
(the one with the highest minimal value, $c_i$) \emph{dies} at the advantage of the oldest
(the one with the lowest minimal value). Critical points $c_i$ and $c_j$ 
form a \emph{persistence pair}.


A classical representation of 
the persistence diagram $\persistentDiagram{f}$ embeds \julien{each} pair 
$(c_i, c_j)$ \julien{as a point}
in the 2D plane 
\julien{at coordinate $\big(f(c_i), f(c_j)\big)$}.
The height of the pair
$P(c_i, c_j) = |f(c_j) - f(c_i)|$ is called the \emph{persistence} and denotes the
life-span of the topological feature created at $c_i$ and destroyed at $c_j$.
In \julien{3D}, the persistence of pairs linking critical points of index
$(0,1)$, $(2, 3)$ and $(1,2)$ respectively denotes the life-span of
connected components, voids and non-collapsible cycles of $\sub{i}$.
\julienLDAV{In the following, we will focus on $(1, 2)$ persistence pairs 
(involving maxima).}

The interest of this visual representation in practice is that 
it quickly hints at the distribution and relative importance of critical points.
Small oscillations due to noise in the input data are typically represented by
pairs with low persistence, in the vicinity of the diagonal.
In contrast, the most prominent topological features are associated with large vertical bars
(\figref{fig:fig1}, b).

Persistence diagrams are used in many applications,
for instance as a visual help for
interactively tuning simplification thresholds in multi-scale segmentation 
tasks, either based on the Reeb graph \cite{reeb46, carr04,
pascucci07, tierny_vis09, gueunet_ldav16,
tierny_vis16} or the Morse-Smale complex
\cite{gyulassy_vis08, gyulassy_vis14, robins11}.

\subsection{\julien{Metrics between Persistence diagrams}}
\label{sec_metric}

Metrics have been defined to evaluate the distance between
scalar fields $f, g : \domain \rightarrow \range$.
The $L^p$-norm $||f - g||_p$ is a classical example.

In the context of \julien{TDA}, multiple metrics
\cite{cohen-steiner05, ChazalCGGO09} have been introduced to
compare \julien{two} persistence diagrams \julien{$\persistentDiagram{f}$ and 
$\persistentDiagram{g}$}.

Critical point pairs in persistence diagrams can be associated with 
a point-wise distance, noted $d_p$ inspired by the $L^p$-norm.
Given two persistence pairs $a = (a_x, a_y) \in \persistentDiagram{f}$ and $b = 
(b_x, b_y) \in
\persistentDiagram{g}$, $d_p$ 
\julien{can be defined as:}
\vspace{-2ex}
\begin{equation}
\begin{split}
  d_p (a, b) &= \big(|a_x - b_x|^p + |a_y - b_y|^p\big)^{1/p} \\
\end{split}
  \label{eq:dnu}
\end{equation}
\vspace{-3ex}

\begin{figure}
 \centering
 \vspace{-2ex}
 \includegraphics[width=\columnwidth]{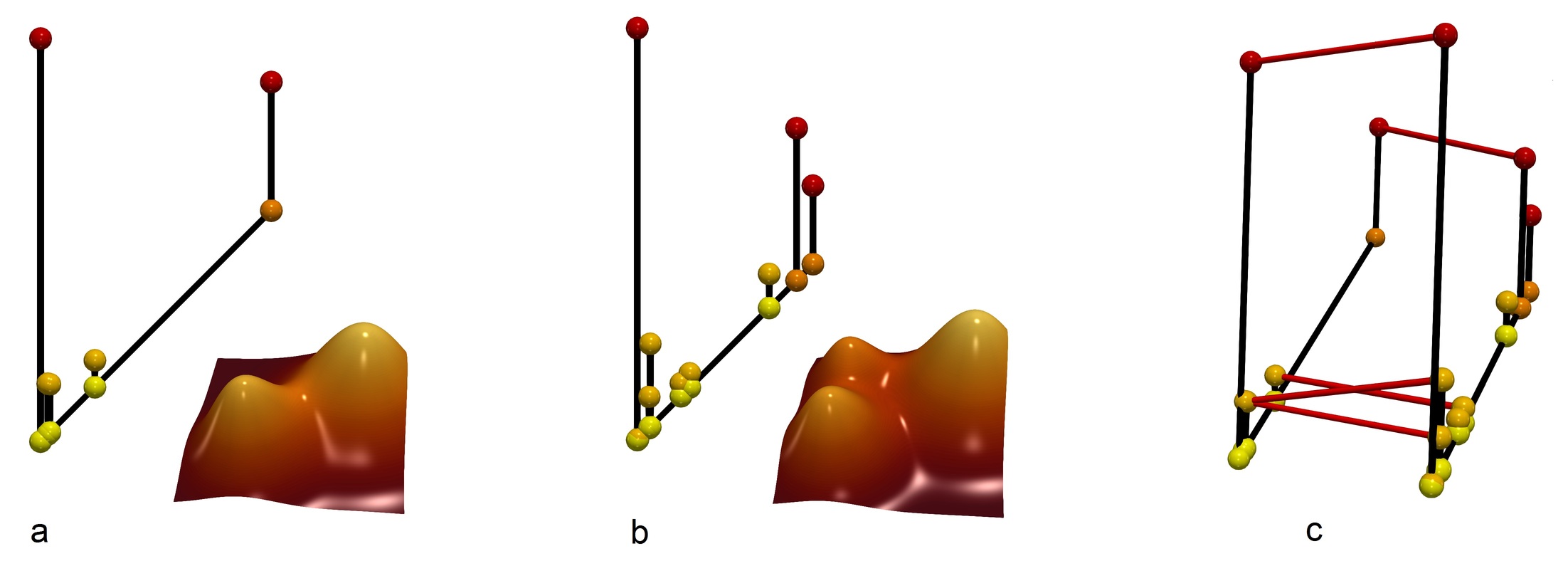}
 \vcaption{\maximet{Persistence diagrams of two distinct 2D 
   scalar fields (a, b), \julienLDAV{matched} 
   by the Wasserstein metric (c, matching pairs are linked 
   with red segments). The third hill of (b)\maximej{, 
   captured by the rightmost persistence pair,} is 
   discarded by the matching. }
 } 
 \label{fig:fig2}
 \vspace{-2ex}
\end{figure}

\noindent
The \emph{Wasserstein} distance \cite{monge81, Kantorovich42},
noted $\wasserstein{p}$, between persistence diagrams
$\persistentDiagram{f}$ and $\persistentDiagram{g}$ 
\julien{can then be defined as:}
\vspace{-1.5ex}
\begin{eqnarray}
  \wasserstein{p} \big(\persistentDiagram{f}, \persistentDiagram{g}\big) =
    \min_{\phi \in \Phi}
      \bigg(
       \sum_{a \in \persistentDiagram{f}}
         d_p
          \big(
            a, \phi(a)
          \big)^p
      \bigg)^{1/p}
      \label{eq:wass}
\end{eqnarray}
\vspace{-2ex}

\noindent
where $\Phi$ is the set of all possible assignments $\phi$ mapping
each 
\julien{persistence pair}
$a \in \persistentDiagram{f}$ to
\julien{a persistence pair}
$b \in \persistentDiagram{g}$ 
\julien{with identical critical indices $\Index$}
or to
\julienWed{its diagonal projection}, noted $\text{diag}(a)$
-- which corresponds to the removal of the corresponding feature
from the assignment, with a cost \maximed{$d_p(a,\text{diag}(a))$}. 
\maximet{It is illustrated in \figref{fig:fig2}.} 
In practice, the Wasserstein distance is computed by solving a 
variant of the assignment problem \cite{munkres, Morozov:Dionysus, kerber17, 
soler2018lifted}.

\julienWed{In the applications,}
this point-wise distance $d_p$ can be fine-tuned 
\julien{ to better account for the layout of critical points}
\julien{in the geometrical domain $\domain$, as done}  in 
applications such as feature tracking \cite{soler2018lifted}\julienLDAV{, 
resulting in the following \emph{lifted} point-wise distance:}

\vspace{-3ex}
\begin{equation}
\widehat{d}_{p}(a,b) =
  (\alpha_x|a_x - b_x|^p + \alpha_y|a_y - b_y|^p +
  \beta_x\delta_x^{p} + \beta_y\delta_y^{p})
  ^{1/p} \\
\label{eq:lifted}
\end{equation}
\vspace{-2ex}


\noindent
\julienLDAV{with}
\julien{$\delta_x = |x_a - x_b|$ and $\delta_y = |y_a - y_b|$ in the 2D 
case, \julienLDAV{and} where $(x_a, y_a)$ stands for the coordinates of the 
extremum of the 
persistence pair $a$ in the geometrical domain $\domain$.}
\maximed{Coordinates are taken at extrema rather than saddles
or midpoints because extrema usually bear more meaning \julien{in the 
applications},
but this \julien{can} be adapted to the specificity of applicative cases.}


\maximed{\julien{In the above equation, t}he coefficients 
$\alpha_x$, $\alpha_y$, $\beta_x$ and 
$\beta_y$
need to be properly tuned \julien{for the target application}.}
\julien{A geometrically lifted version of the Wasserstein distance, noted 
$\liftedWasserstein{p}$, can then be introduced as:}
\vspace{-2ex}
\begin{eqnarray}
  \liftedWasserstein{p}
  \big(\persistentDiagram{f}, \persistentDiagram{g}\big) =
    \min_{\phi \in \Phi}
      \bigg(
       \sum_{a \in \persistentDiagram{f}}
         \widehat{d}_p
          \big(
            a, \phi(a)
          \big)^p
      \bigg)^{1/p}
      \label{eq:lifted_wass}
\end{eqnarray}
\vspace{-2ex}



\section{\julien{Analysis framework}}

\maxime{
This section describes the problem of 
representing viscous fingers appearing 
\julien{in time-varying saturation fields,} 
comparing them across  simulations, and our approach for addressing
this problem. 
\julien{In the following, we will note each time step of the 
reference ground-truth acquisition $A_t$ and each time step of a simulation 
run $S_t$.
Then, the goal of our framework is to efficiently compute relevant
similarity measures,
to rank simulation runs in order of increasing distance to the 
acquisition, \maximej{so as} to present to the experts the most plausible 
simulations for 
further inspection \maximej{(\figref{fig:teaser})}.}}

\maxime{\maximed{
\julienLDAV{
Note that, in the rest of the paper, we will consider only one ground-truth 
acquisition data set, as the acquisition process (further detailed in 
\secref{sub:protocol}) is long (several months) and involves expensive 
instrumentations (the process requires the acquisition machinery to operate in 
a high pressure environment to replicate the reservoir conditions). Because of 
the rarity of such acquired data, }
the number of \emph{available} acquired time steps 
$A_t$
is in practice significantly lower than the number of
simulated time steps $S_t$. The simulator is \julien{thus} set \maximej{up} to output
additional 
time steps \maximej{corresponding to} a specific set of volumes of injected water,
\maximej{which were} recorded \julien{for each time step $A_t$}. 
This physical criterion allows \julienLDAV{us} to reliably
match \julien{in time acquired and simulated time steps. \julienLDAV{The 
correspondence between simulation and acquisition timesteps is then given 
and reliable.}}}
}

\subsection{Feature representation}
\label{sec:features}

\begin{figure}[b]
 \centering
 \includegraphics[width=\columnwidth]{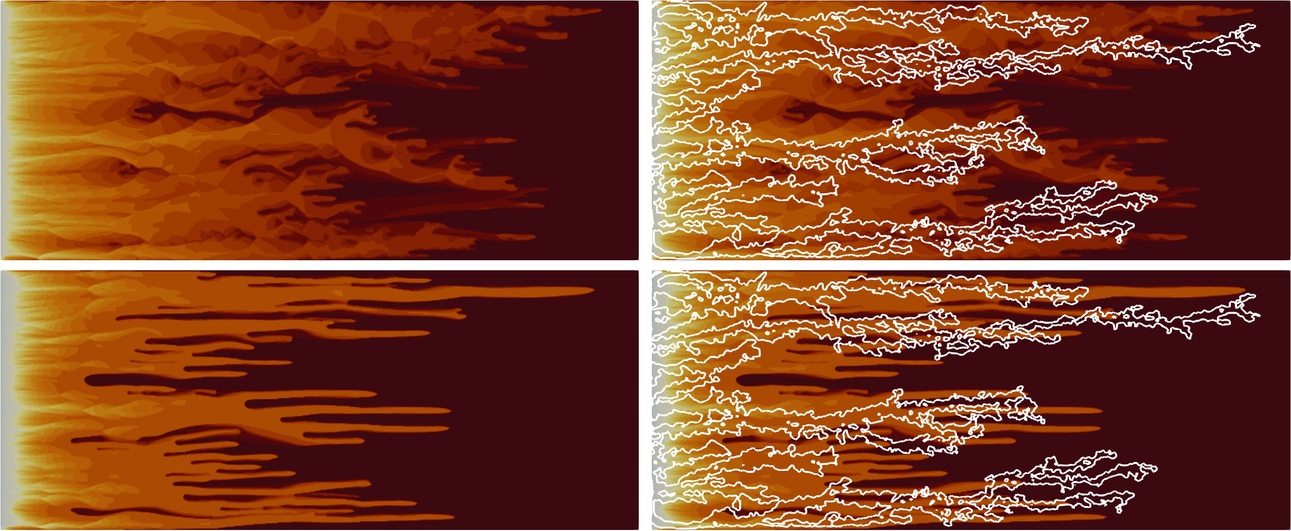}
 \vcaption{\maximet{Late time-steps of two Darcy-type simulation runs 
   launched with different model parameters (left column).
   The ground truth obtained with X-rays is contoured 
   in white (right column\maximej{, superimposed}). Runs exhibit a very chaotic 
   finger geometry.} } 
 \label{fig:fig3}
 \vspace{-2ex}
\end{figure}

\maxime{As \julien{discussed} in the introduction, trying to reproduce 
the viscous fingering phenomenon with Darcy-type simulation software
is \julien{very} challenging because 
the \julien{fingering}
geometry \julien{greatly varies when \maximej{one modifies} input parameters, even 
slightly.}
\julien{In particular, the}
input parameters \julien{considered here}
are \julien{the} relative permeabilities $kr_i$.} \julien{
When comparing a simulation to 
an acquired ground-truth, this great geometrical 
variability 
challenges
traditional image based distances, either point-wise based 
($L_2$ norm) or morphing based \cite{Cuturi13}.}
\julien{Moreover, the raw geometry of the viscous fingers can be insufficient 
in practice to identify all plausible simulations. Indeed, two geometrically 
 different simulations can be deemed equally plausible by the experts 
if they share more abstract similarities, involving the number of fingers, 
their prominence and their progress in the porous medium. Thus, a proper 
feature representation, capable of abstracting these informations, is required 
to correctly represent the viscous fingering.}
\maxime{\figref{fig:fig3} illustrates the extent to which the geometry of fingers
may vary across simulations \julien{and how clearly distinct simulations can 
be judged as equally plausible by the experts}.}

\maxime{The water saturation scalar field allows \julienLDAV{one} to visually 
identify
fingers, because they form a clear, sharp frontier with the background
(as the geometric domain was initially filled with oil).
The first step for identifying fingers then consists in
\julien{extracting a sub-level set $\sub{w}$ of water saturation, for an 
isovalue $w$ chosen properly,}
to extract the geometric domain $\domain$  where fingers are effectively 
present. 
\julienLDAV{In our use case, based on discussions with experts, we set in 
practice this isovalue parameter once for all to $0.12$.}
\julien{Let $\mathcal{F} = \sub{w}$} be that sub-part 
of $\domain$.
The same workflow can be applied on acquired X-ray images.}

\julien{To compare a simulation to an acquired ground-truth, a naive strategy 
consists in estimating}
\maxime{overlaps between 
\julien{the sub-level sets of saturation of the simulation and the acquisition, 
for a given time-step,}
and use \julien{the area of such an overlap} as a measure of likeliness.
However, this purely geometric approach appears to be inadequate \julien{in 
practice}
due to the \julien{important} variability in the number and shape of fingers, 
which then would not be accounted for \julienLDAV{(see \figref{fig:fig3})}.}

\begin{figure}
 \centering
 \vspace{-2ex}
 \includegraphics[width=\columnwidth]{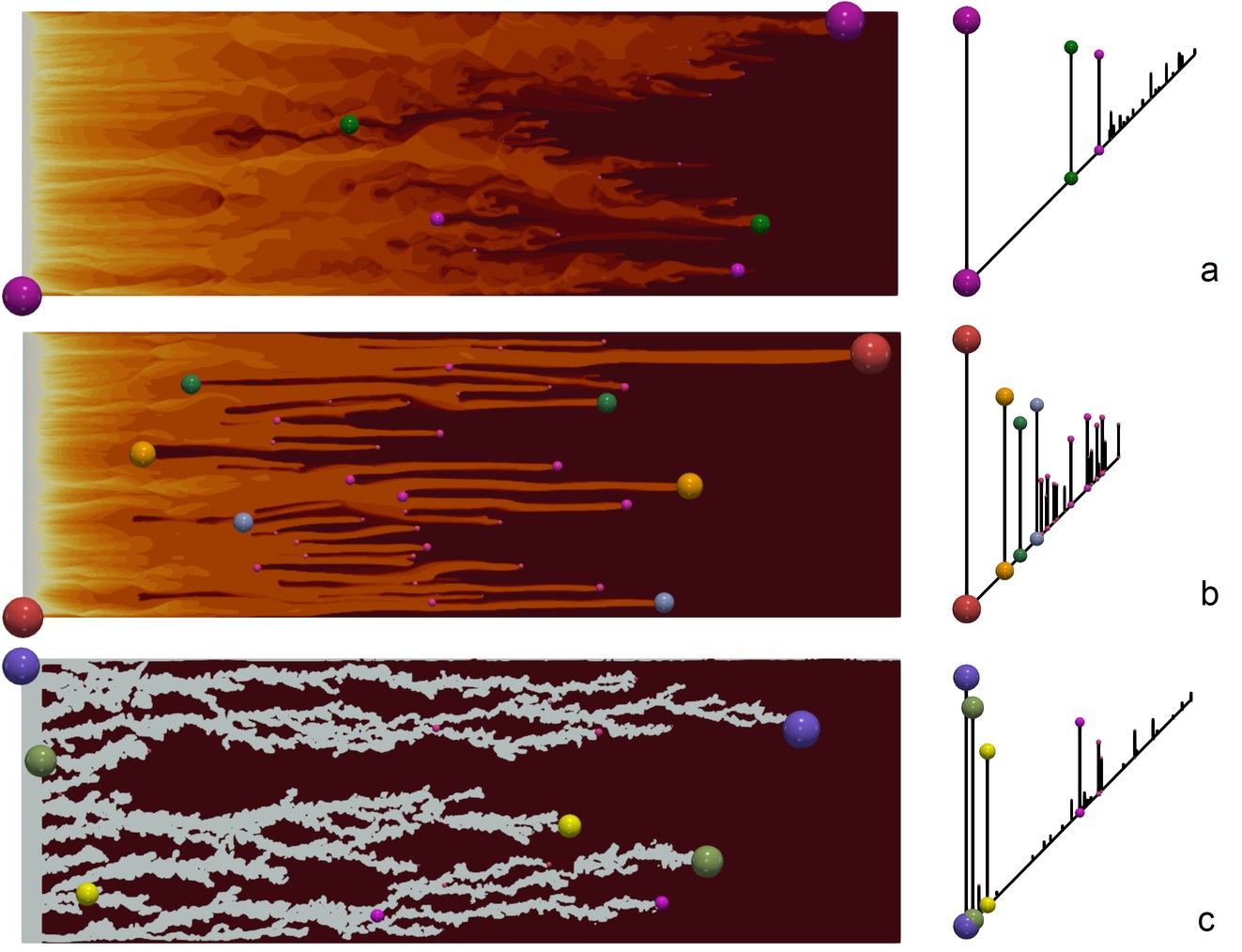}
 \vcaption{\maximet{Simulated time-steps (left column, a and b)
 and the matching ground truth image (left column, c). 
 Critical points are represented with spheres, and
 the corresponding persistence diagram is shown on the right.
 As every critical point belongs to only one persistence pair,
 the color of spheres encodes their persistence pair; 
 and their diameter encodes its height (the larger, 
 the higher the persistence). The most important fingers can 
 clearly be identified by looking at the most persistent 
 pairs in diagrams (right column). For instance, we can see that
 the three most important fingers in the acquisition are the purple, 
 green and yellow ones. }
 } 
 \vspace{-2ex}
 \label{fig:fig4}
\end{figure}

\maxime{A natural way of characterizing fingers while taking
their shape into consideration is to provide 
\julien{$\mathcal{F}$}
with a 
\julien{\emph{descriptive}}
scalar field, for instance a geodesic distance
from the injection point. Here, the injection point is \julien{the}
left boundary of the domain, so the scalar field can simply be
the $x$ geometrical coordinate.
Local maxima of this new scalar field would then correspond to 
the tips of viscous fingers, and saddles to \julien{valleys} between fingers.}
\julien{Since they correspond to finger tips, maxima of the $x$ 
geometrical coordinate provide a useful information to represent the progress of 
each finger in the porous medium. Moreover, in this setting, the persistence 
of the 
pair involving each maximum directly represents the length of the 
corresponding finger, which can be used as a reliable measure of importance 
given this application, to distinguish the main fingers from noise.}
\julien{The persistence diagram directly captures this information,}
\maxime{in a robust and hierarchical setting.
\figref{fig:fig4} illustrates the correspondence between fingers in the domain and 
pairs of critical points in persistence diagrams.
In this context, persistence diagrams seem to be a promising
feature representation for viscous fingers, \julien{since they efficiently 
describe their number, progress through the porous medium as well as their 
prominence.}}

\subsection{Metric\maxime{s} between time-varying persistence diagrams}
\label{sec:metrics}

\maxime{Considering \maximej{that} viscous fingers are captured by persistence diagrams,
computing the 
\julienWed{similarity} of a simulation with respect to a ground truth
would require, in a first step, to compute distances between persistence diagrams.
As outlined in the introduction, metrics have been introduced for this purpose, notably 
the (2-)Wasserstein distance in the \emph{birth-death} space,
\julien{noted $\wasserstein{2}$ (\eqrref{eq:wass}).}}

\begin{figure*}[ht]
 \centering
 \vspace{-2ex}
 \includegraphics[width=\linewidth]{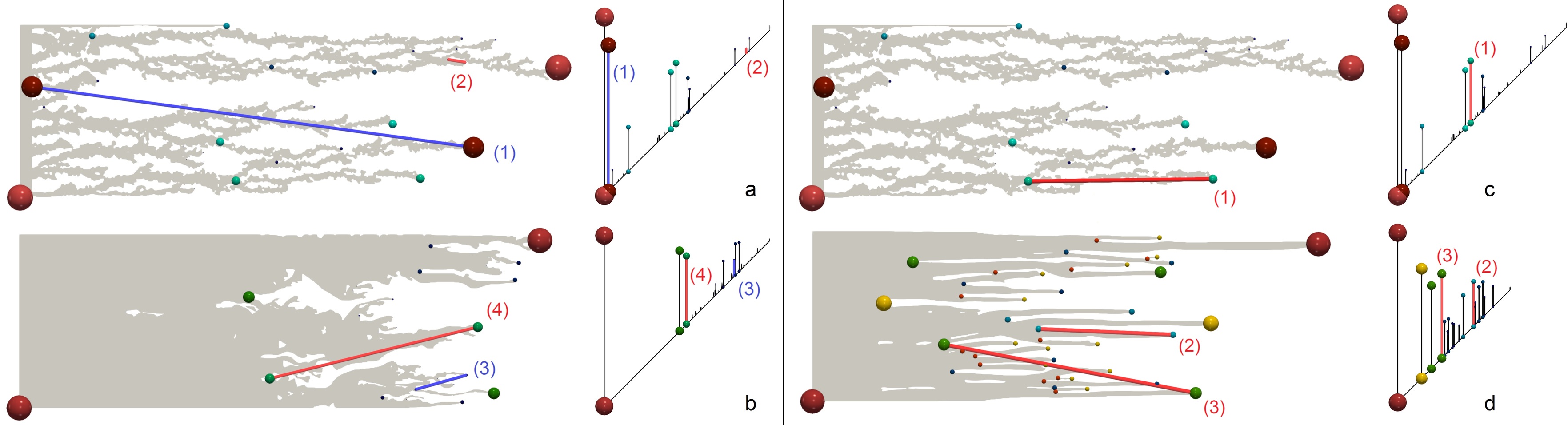}
 \vcaption{\maximet{Limitations of matching methods based on 
   geometry only (a, b) and matching methods based on persistence 
   only (c, d). As the Earth mover's distance (left) only considers 
   the geometrical location of extrema, it can incorrectly associate 
   critical points belonging to unrelated fingers. For instance, the 
   large pair in the acquisition, represented with a blue segment (a, (1)), 
   is matched to a pair with low persistence (b, (3)) \maximej{because their maxima are
   geometrically close}; 
   and the large red finger in the middle of a simulation (b, (4)) is matched 
   to a small protrusion (a, (2)) attached to the largest finger in the acquired image. 
   On the right, the reference metric for matching persistence diagrams, 
   the 2-Wasserstein metric, is shown to associate the bottom finger 
   in the acquisition (c, (1)) to a finger in the middle of a simulation (d, (2)),
   because their persistence is \maximej{comparable}.
   Taking both geometry and persistence into account, a \emph{lifted} 
   version of the Wasserstein metric 
   associates (c, (1)) to (d, (3)), which is farther away in terms of 
   persistence, but has the nearest maximum, and is qualitatively the best match.
 } } 
 \label{fig:fig5}
\end{figure*}

\maxime{A drawback of $\wasserstein{2}$ is that it does not take \julien{into 
consideration} geometrical information,
other than coming from the birth-death space. \figref{fig:fig5} 
illustrates this limitation. To avoid this problem, a \emph{lifted} adaptation 
of $\wasserstein{2}$\julien{, noted $\liftedWasserstein{2}$,}
including geometrical components \julien{can be considered 
(\eqrref{eq:lifted_wass}).}
It is subject to input parameters indicating the importance given to each 
geometrical component.
\julien{Note that, the \emph{Earth mover's distance} \cite{levina01}, 
noted $EMD$, which is an 
alternative of interest too for our application, is a special case of 
$\liftedWasserstein{2}$, for $\alpha_x = \alpha_y = 0$.}
It is similar to $\wasserstein{2}$, but \julien{it only} operates on the 
geometrical space instead of the birth-death space.}
\julien{Thus, the lifted Wasserstein distance $\liftedWasserstein{2}$ can be 
interpreted as a blend between the $\wasserstein{2}$ distances in the 
diagram birth-death space and 
in the geometrical domain.}

\maxime{In practice, an important characteristic of a viscous fingering
simulation run is the moment when the longest finger arrives at the right boundary,
called \emph{breakthrough time}. Correctly predicting this event is essential
because once it is reached, it means a preferential path has been formed, 
allowing water 
to easily flow through, impacting production.
\julienLDAV{Thus}, the position of local maxima (i.e. fingertips) is more 
important
than the position of saddles (i.e. finger branchings).}

\begin{figure}
 \centering
 \includegraphics[width=\columnwidth]{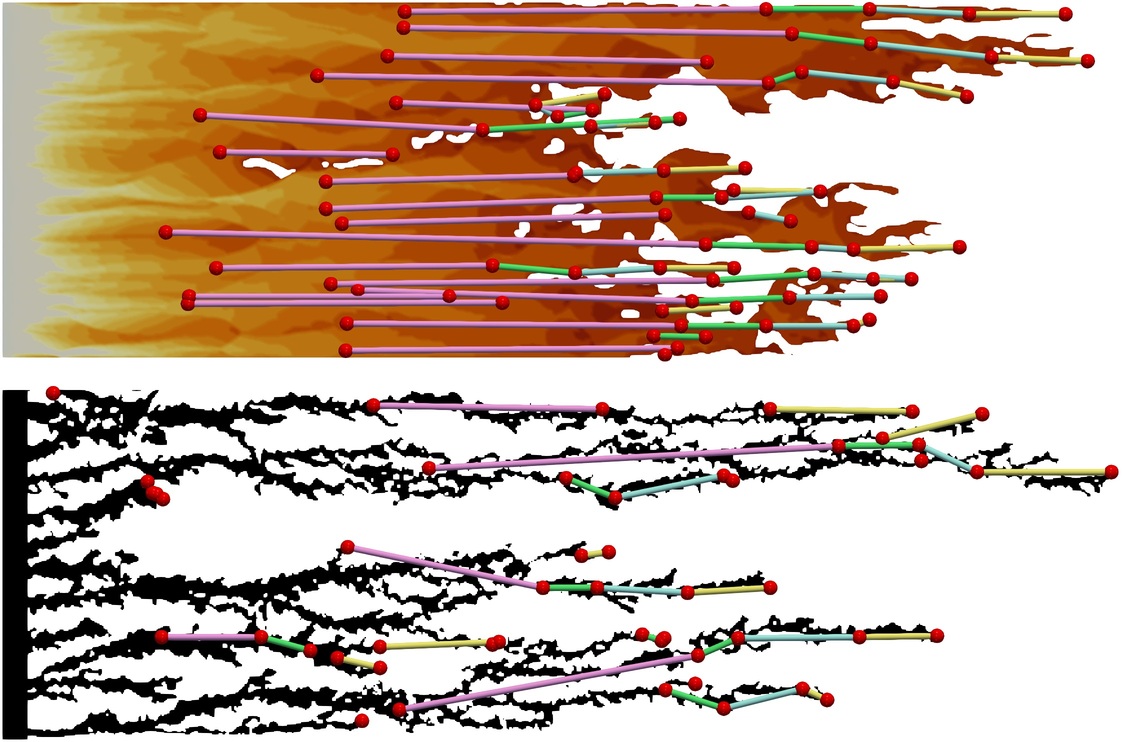}
 \vcaption{\maximet{Critical point trajectories 
 based on optimal matchings. 
 Within a given simulation 
 (or the \maximej{acquisition, bottom}), the geometrical coherence of fingers 
 allows \julienLDAV{us} to use a lifted version (that gives importance to 
 the $y$-coordinate of fingers) of the Wasserstein metric 
 to correctly track the evolution of persistence pairs. 
 Comparing the mean distance \maximej{traveled} by fingers between
 simulations and the acquisition, for each pair of time-steps, 
 is proposed as a \emph{velocity}-aware metric. 
 } } 
 \label{fig:fig6}
\end{figure}

\maxime{
\julien{Then, given a time-step $t$, to compare the}
persistence diagrams coming from a simulation $S_t$ and the 
acquisition $A_t$,
metrics should be more sensitive to the advancement of fingertips, 
then to the global extent of fingers, and lastly to their $y$ location in the domain.
\julien{Thus, at}
this point, we propose to select the following metrics:
\begin{itemize}
\vspace{-1ex}
  \item{The Earth mover's distance for local maxima: $EMD(S_t, A_t)$}
  \vspace{-1ex}
  \item{The 2-Wasserstein distance: $\wasserstein{2}(S_t, A_t)$}
  \vspace{-1ex}
  \item{The 2-Wasserstein distance, lifted to 
  \maximet{include geometrical information (the position of critical points): 
  $\liftedWasserstein{2}(S_t, A_t)$. As in this application, 
  the advancement of fingertips is much more
  important than their vertical position in the domain,
  we only consider the $x$-coordinate of critical points.} 
  \maximet{Thus, lifting coefficients \maximet{(cf. \eqrref{eq:lifted})} are $\beta_x = 10/\gamma$ 
  ($\gamma$ being the extent of the geometrical domain),
  $\beta_y = 0$, and $\alpha_x = \alpha_y = 1/\rho$ ($\rho$ being the 
  range of the scalar function). \julienLDAV{The values of these 
lifting parameters have been adjusted 
empirically
based on discussions with experts.}} }
\vspace{-1ex}
  \end{itemize}}
\maxime{Characterizing the evolution of fingers through time raises the necessity
to integrate these metrics, as they are intended to evaluate the proximity
between persistence diagrams for a single time-step \julienWed{$t$}.
\julien{Thus, to measure the distance from a time-varying simulation $S$ to the 
time-varying acquired ground truth $A$, we introduce time-integrated versions,
based on the $L_2$ norm, 
of the \maximeLDAV{above metrics}:
\begin{itemize}
  \vspace{-2ex}
  \item{$d_{EMD}(S, A) = \Big(\sum_t \big(EMD(S_t, A_t)\big)^2\Big)^{1/2}$}
  \vspace{-2ex}
  \item{$d_{\wasserstein{2}}(S, A) = \Big(\sum_t \big(\wasserstein{2}(S_t,
A_t)\big)^2\Big)^{1/2}$}
\vspace{-2ex}
  \item{$d_{\liftedWasserstein{2}}(S, A) = \Big(\sum_t \big(
\liftedWasserstein{2}(S_t, 
A_t)\big)^2\Big)^{1/2}$}
\vspace{-1ex}
\end{itemize}
}}
%
\maximed{As suggested by the experts, the displacement speed of the saturation
front is key to predicting breakthrough time. They suggested to 
match in priority simulations which display compatible fronts in terms of velocity
during the experiment. Given fingers are captured by persistence diagrams, 
a possibility for appreciating their evolution with respect to that suggestion
would be to compute the sequence of distances between diagrams in successive time steps.
In other words, for each couple of consecutive time steps
\julien{$t$ and $t+1$}, compute \julien{a distance between $S_t$ and $S_{t+1}$ 
(\autoref{sec_metric}), and \julienWed{integrate} for all time steps.}
Here the chaotic behavior displayed by fingers when input simulation parameters change
need not be taken into account: we are considering a unique simulation run, which has
temporal coherence, therefore it is easier to choose a fitting metric.
As \julienLDAV{shown} in \figref{fig:fig6}, a working solution is the 
2-Wasserstein distance, 
lifted to give more importance to the $y$ coordinate of maxima: 
\maximet{$\otherLiftedWasserstein{2}(S_t, S_{t+1})$}\maximej{, with 
lifting coefficients $\beta_x = 0$, $\beta_y = 10/\gamma$, $\alpha_x = \alpha_y = 1/\rho$ 
($\gamma$ is the geometrical extent; $\rho$ is the scalar range)}.
Because of the variability in the number of fingers, however, 
considering the \maximet{difference} of traveled distances alone could be problematic,
for many little fingers going slow could compare close to few fast fingers.
We then consider the mean traveled distance per finger.
\maximet{Thus, if $n_{A_t}$ (resp. $n_{S_t}$) denotes the number of fingers in the acquisition
(resp. simulation) at time-step $t$,} 
we propose to evaluate the velocity-oriented difference by:}
\maximet{
\vspace{-2ex}
\begin{itemize}
  \item{$d_{\otherLiftedWassersteinIndex{2}}(S, A) = \Big(\sum_t \big(
  \frac{1}{n_{S_t}} \otherLiftedWasserstein{2}(S_t, S_{t+1}) - 
  \frac{1}{n_{A_t}} \otherLiftedWasserstein{2}(A_t, A_{t+1})\big)^2\Big)^{1/2} $}
\end{itemize}
\vspace{-2ex}
}

%

\julien{Then, given an ensemble of time-varying viscous fingering simulations, 
each run $S$ can be compared to the reference acquired ground-truth $A$ and 
runs can be ranked in increasing order of distance to $A$ and presented to 
the experts for further visual inspection.}


\subsection{In-situ deployment}
\label{sec:insitu}

\maxime{Doing feature extraction and comparisons can be problematic
for very large numbers of simulations, in terms of data movement.
Fortunately, computing the metrics we just presented does not
require to have all time-steps available at once, and hence may be 
done in a progressive fashion.}
\maxime{We propose, within our framework, to implement the
computation of metrics comparing 
\julien{the acquired reference to the simulation}
\emph{in-situ}, that is, without storing time-steps to the disk first.
Precomputed persistence diagrams for 
\julien{the acquisition}
are first \julien{loaded in memory}.
Whenever the simulation attains a time for which
there is a \maximed{corresponding} 
\julien{acquisition time step,}
the saturation scalar field is passed 
to our analysis pipeline, which applies a threshold, \julien{extracts} the
persistence diagram, 
\julien{and computes}
the \julien{per time step} distance to the acquisition diagram \julien{(for 
instance 
$\liftedWasserstein{2}(S_t, A_t)$) }. The distance 
can then be accumulated
as the simulation unfolds. 
\maximed{The \emph{in-situ} application of our pipeline is optional: 
time-steps can still be saved to the disk and the pipeline applied
\emph{post-mortem} if desired. } }


\subsection{Visual interface}
\label{sec:visualInterface}

\begin{figure}[ht]
 \centering
 \vspace{-2ex}
 \includegraphics[width=\columnwidth]{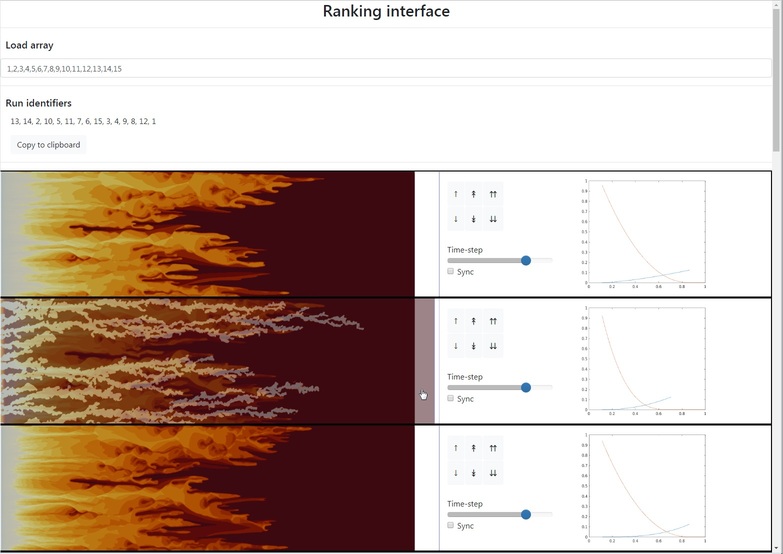}
 \vcaption{\maximet{Lightweight web interface for exploring and modifying
 simulation rankings.
 An \maximej{ordered} list of runs can be loaded as 
 an input. Time-steps of runs are then displayed on the left pane;
 they can be hovered with the mouse to be compared with the 
 matching acquired image. A slider allows to select the time-step 
 to compare. Users can edit the ranking with swapping buttons.
 For each run, $kr$ curves 
 (input parameters of the simulation model) are \maximej{displayed} on the right.} } 
 \label{fig:fig9}
\end{figure}

\maxime{Each metric previously mentioned naturally produces
a ranking of simulations, from the most to the less 
\julien{plausible}
ones.
We propose a way to visually inspect those rankings 
with a lightweight HTML+Javascript application, as illustrated 
in \figref{fig:fig9}.
\julienLDAV{Note that we} 
use the same interface \julienLDAV{for two tasks},
to allow experts to \julienLDAV{\emph{(i)} visually explore the rankings 
generated by our framework and to \emph{(ii)} produce a ground-truth 
reference ranking (for the quantitative evaluation, \secref{sub:protocol}).}
\julien{This visual interface offers linked views of the saturation scalar 
fields, to visually compare simulations runs, for a given time step $t$ which 
can be interactively selected. If needed (in particular to generate a 
reference ranking, cf. \secref{sub:protocol}), the experts can interactively 
modify the suggested ranking by displacing a selected run up or down the 
ranking, either by unit or long jumps (typically 
skipping \maximet{$10$ or $50$} positions\julienLDAV{, useful for the a priori
reference ranking}).}}

\section{\julien{Case study}}

\maxime{This section exposes our experimental setting,
details a complete viscous fingering use case and 
summarizes the results of our approach in terms of performance and quality,
compared to classical methods.}


\subsection{Experimental protocol}
\label{sub:protocol}

\maxime{
The behavior of a \maximej{slab}, initially
filled with oil \maximed{and water at connate water saturation, }
then subject to a water
injection in reservoir conditions 
is captured through X-rays\maximed{: X-ray images are processed in order to be converted to
maps of the fluid saturations within the \maximej{slab}.}
\maximej{2D s}imulations are then launched with varying input parameters
in order to match 
\maximed{the simulation results to the experimental measurements
and to the fluid saturation maps derived from X-ray images.} 
The resemblance of fingers can be taken into account manually by experts,
involving an interpretation of X-rays and an assessment of likeliness
according to their expertise.
A reference ranking of simulations is then 
\julien{produced by the experts with the help of our visual interface 
(\secref{sec:visualInterface}),}
and is compared to the rankings generated by the 
\julien{metrics proposed in our framework (\secref{sec:metrics}).}
The performance and quality of our approach are then evaluated.}

\begin{figure}
 \centering
 \vspace{-2ex}
 \includegraphics[width=\columnwidth]{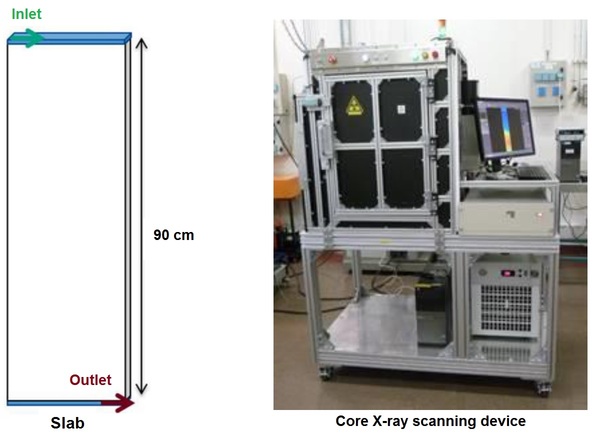}
 \vcaption{\maximet{Schematic view of the slab 
 used for the acquisition (left, experimental 
 protocol described in \cite{de2018numerical}). 
 It is disposed vertically during the capture.
 On the right, a typical X-ray scanning device 
 for imaging flow in porous media is shown.} } 
 \vspace{-2ex}
 \label{fig:fig10}
\end{figure}

\noindent
\textbf{Acquisition: }
\maxime{The detailed experimental setup is that of \cite{de2018numerical, 
fabbri2015comparison},
further described in \cite{skauge2014polymer, skauge2009experimental, skauge20122}.
We consider slabs of Bentheimer sandstone 
(30$\times$90$\times$2.45 cm),
with porosity of approximately 23\% and
\maximed{absolute} permeability of 2.5 Darcy (when $S_w=1$).
The slab is coated with two epoxy layers. Three grooves are cut into the 
first layer on the extreme faces, and connected to injection and 
production rails. It is mounted vertically in a 2D X-ray scanning rig (\figref{fig:fig10}).}

\noindent
\maxime{Slabs first undergo cleaning and calibration processes.
A tracer test validates the homogeneous behavior of the slab, then 
oil is injected to reach initial conditions. The system is then aged at
50\degree C and ambient pressure for a month,
to get closer to field conditions.}
%
\maxime{
The water injection rate is kept constant at 3 cm\textsuperscript{3}/h,
which corresponds to the velocity in fields far from wells.
One of the fluids is doped with an X-ray absorbing chemical for increasing the contrast.
The scanner is equipped with an X-ray source (40 to 60 kV at maximum 0.4 
mA) and 
a camera capturing a slice of 0.5$\times$11.5 cm. The camera
moves in horizontal rows along the slab. 
\julienLDAV{A scan for a 30$\times$30 cm image}
takes 4-5 min, during which the fluid has moved by about 0.1 to
0.2 mm.}
\noindent
\maxime{The captured images, which are noisy and exhibit severe vertical
and horizontal artifacts, are filtered \cite{skauge20122}, 
and manually segmented \julien{by an expert} (\figref{fig:fig11}) to differentiate fingers from the 
background, \julien{hence} forming a reference finger 
geometry \julien{$\mathcal{F}_A$}.}

\begin{wrapfigure}{r}{0.5\linewidth}
  \begin{center}
    \vspace{-7ex}
    \includegraphics[width=\linewidth]{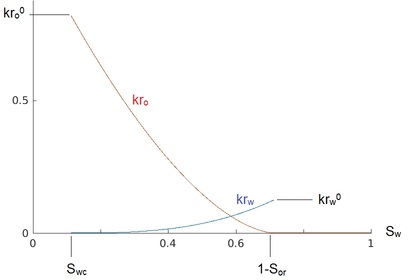}
  \end{center}
  \vspace{-2ex}
  \vcaption{\maximet{Typical relative permeability curves: 
 for oil (red) and water (blue). In this graph, the 
 \julienLDAV{X-axis}
 is 
 water saturation $S_w$. Intuitively, it represents the extent to which 
 the flow of a phase (say the flow of oil) is inhibited by the 
 presence of another (say the presence of water).} } 
 \label{fig:fig12}
\end{wrapfigure}

\begin{figure}
 \centering
 \includegraphics[width=\columnwidth]{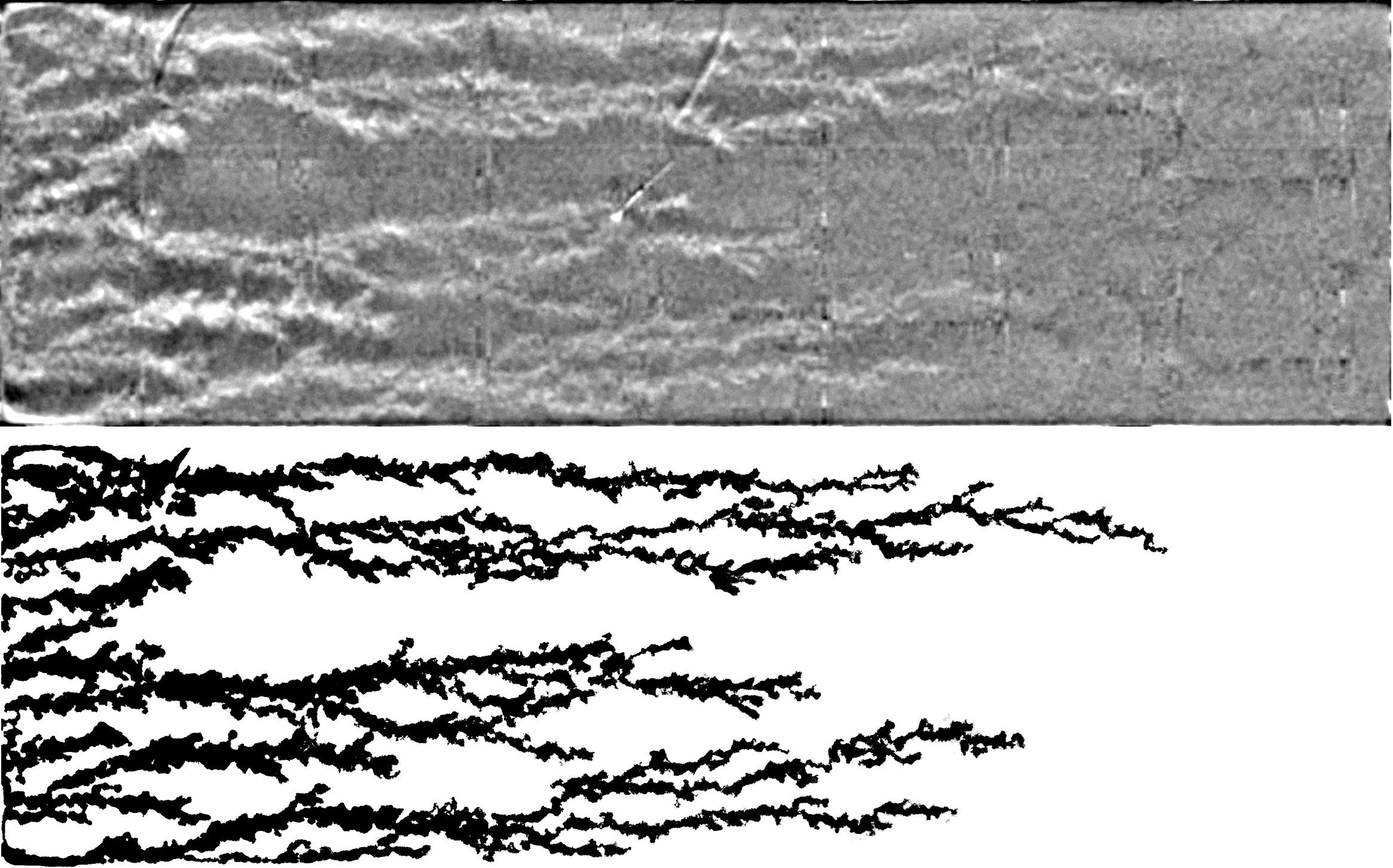}
 \vcaption{\maximet{De-noised X-ray capture (top) and 
 segmented fingers (bottom). Fingers were manually detoured
 by experts.}
 } 
 \vspace{-2ex}
 \label{fig:fig11}
\end{figure}

\noindent
\textbf{Simulations: }
\maxime{The input parameters of \maximej{the 2D} simulations are relative 
permeabilities \maximed{($kr_w$ and $kr_o$)}.
In our model, they are a function of water saturation $S_w$.
We consider relative permeabilities in the form of simple Corey curves
(\figref{fig:fig12}, \eqrref{eq:corey}, \cite{brooks1964hydraulic}),
subject to parameters $kr_{o}^0$ 
(oil relative permeability \maximed{endpoint}),
$S_{or}$ (residual oil saturation), and power law exponents $n_c$ and $n_w$.
Other quantities like $S_{wc}$ (connate water saturation) and $kr_w^0$ (water 
relative permeability endpoint) are determined \maximed{by measurement.} } 

\vspace{-2ex}
\begin{equation}
  \begin{cases}
    kr_o(S_w) = kr_o^0 \times
          (\frac{1-S_w-S_{or}}{1-S_{or}-S_{wc}})^{n_c}\\[1em]
    kr_w(S_w) = kr_w^0 \times 
          (\frac{S_w-S_{wc}}{1-S_{or}-S_{wc}})^{n_w} \\
  \end{cases}
  \label{eq:corey}
\end{equation}

\noindent
\setlength{\columnsep}{10pt}%
\maxime{The parameters of these curves\maximej{,
$kr_o^0$, $S_{or}$, $n_c$ and $n_w$,} 
were randomly sampled and selected using the 
\julien{algorithm by}
Wootton, Sergent, Phan-Tan-Luu \cite{santiago2012construction}
to ensure a good initial covering of the space.
The geometrical domain is discretized on a regular grid of 290$\times$890 blocks.
200 runs were launched on 400 simulation nodes (2 MPI ranks per run),
then time-steps for which there was a corresponding 
X-ray image were saved \maximed{(8 available segmented images)}.}

~

\noindent
\textbf{Expert ground truth: }
\maxime{Images of the water saturation field were captured for
each simulation at available X-ray time-steps, for experts to manually 
form a reference ranking.
During this process, experts would quickly discard runs
\julienLDAV{deemed}
too far from the  X-ray image (because fingers are advancing too 
slow,
too fast, or in a too diffusive fashion).
Then, they would closely look at the shape and advancement of fingers when 
comparing two close runs.
We 
use our lightweight web \julien{based visual} interface 
\julien{(\secref{sec:visualInterface}, \julienLDAV{\figref{fig:fig9}})},
to alleviate this tedious process.
\maximed{Note that images from all simulations were necessary for
the experts to form the reference ranking, so the corresponding 
simulated time-steps were saved to the disk. Once this reference is
formed, later analyses can be done \emph{in-situ}. }
}

\subsection{\julienLDAV{In-situ performance}}

\maxime{In this \julien{section}, we evaluate the \julien{quantitative gains} 
of using our approach 
\emph{in-situ} \julien{(\secref{sec:insitu})}, in terms of time and storage.}
\julien{\tabref{tab:performance} provides }
\maxime{a CPU time comparison
of \julien{our} analysis pipeline based on \julien{the lifted} 2-Wasserstein 
metric 
(\secref{sec:metrics}), for the two different strategies: 
\julien{\emph{(i) in-situ}, where the analysis is run on the fly during the 
simulation and without data storage and \emph{(ii) post-mortem}, where selected 
time steps are stored to disk to be analyzed after the simulation has finished}.
Persistence diagrams are computed using the 
\julien{algorithm}
by \julien{Gueunet et al.} \cite{gueunet_ldav17}, and Wasserstein distances are 
computed using the exact 
\julien{approach}
by \julien{Soler et al.} \cite{soler2018lifted},
both available in the Topology ToolKit (TTK, \cite{ttk}). 
The \emph{in-situ} implementation is based on
 Catalyst \cite{ayachit2015paraview}\julien{, which is called by the 
simulation code at selected time steps to run a python script 
instantiating our analysis pipeline.}}

\begin{table}
    \centering
    \caption{Time performance comparison (CPU time in seconds)\julien{, for a 
single time-step,}
    between the \emph{in-situ} implementation (everything is computed during
    the simulation using the local CPU's memory)
    and the \emph{post-mortem} implementation (the scalar fields are
    analyzed and compared in a post-processing stage).}
    \label{tab:performance}
    \scalebox{0.725}{
    \centering
    \begin{tabular}{|l|r|r|ll|}
    \hline
    \rule{0pt}{1.0\normalbaselineskip}Step
    & \multicolumn{2}{c|}{CPU time (s)}
    & \multicolumn{2}{c|}{Detailed CPU time (s)} \\
    & \multicolumn{1}{c}{\emph{in-situ}} & \emph{post-mortem} 
    & \multicolumn{2}{c|}{} \\
    \hline
    \rule{0pt}{1.0\normalbaselineskip}Simulation iteration 
                        & 3.096 & 3.096   & & \\
    \hline
    Time step storage   &       & 0.076   & & \\
    \hdashline
    Catalyst analysis   & 1.111 &         & 0.063  & Persistence diagram \\
                        &       &         & 0.002  & Distance  \\
                        &       &         & 0.001- & \maximed{Distance storage} \\
                        &       &         & 1.046  & Catalyst overhead \\
    \hdashline
    \rule{0pt}{1.0\normalbaselineskip}Data transfer 
                        &       & 0.021   &        & Lustre to workstation \\
    \hdashline
    Data conversion     &       & 0.246   &       & .unrst to .vtk \\
    \hdashline
    Paraview analysis   &       & \maximed{2.189}  & 0.084 & Persistence diagram \\
                        &       &                  & 0.002 & Distance \\
                        &       &                  & \maximed{2.085} & Paraview overhead \\
    \hline
    \julien{Analysis time} & \textbf{1.111} & \textbf{\maximed{2.532}} & &\\
    \hline
    \julien{Total processing} & 4.207 & \maximed{5.628} & &\\
    \hline
    \end{tabular}
    }
    \vspace{-2ex}
\end{table}

\maxime{The \julienWed{numbers} are given for a single simulation time-step,
therefore at the finest possible time resolution (about ten thousand time-steps
are required to complete a run).
Figures are averages \maximej{on the time-steps of a typical} run.
\emph{In-situ} computations
are done on a supercomputer 
\maximed{(among the $51^{st}$ of TOP500 Nov. 2018)} 
with Xeon(R) E5-2680v3 processors,
the post-process is done on a local workstation with a Xeon(R) E5-2640v3 processor,
so there is a difference in performance obtained
for computing persistence diagrams.}

\maxime{Ideally, overheads due to different data layouts 
\maximed{and conversions} 
\julien{(line\maximed{s} 
\emph{``Catalyst overhead''} \maximed{ and \emph{``Paraview overhead''}}, 
\tabref{tab:performance})} in 
the simulator and \julien{VTK/ParaView}
would not enter into account (if the simulator \julien{were to}
directly output a \julien{VTK} data array). 
We are left with two unneeded stages in the \emph{post-mortem} approach:
time step writes and data transfer. 
\julien{Selecting}
$8$ time-steps from $200$ simulations, this 
\julien{amounts to}
$155.2$ s. of IO time
versus \julien{approximately} $0.6$ ms. necessary to write \julien{the} $1,600$ 
doubles  in the 
\emph{in-situ} case \julien{(representing the $1,600$ distance estimations)},
which is $260,000$ times faster.}

\maxime{In terms of data storage, the \emph{post-mortem} strategy requires
to store and potentially transfer $3.28$ \maximej{GiB} ($2.1$ \maximej{MiB} per time-step) of data,
versus $12.5$ \maximej{KiB} for $1,600$ doubles \julien{(representing the $1,600$ distance 
estimations)}, which is $275,000$ times lighter.}

\julien{Thus, overall, the in-situ instantiation of our framework reduces 
 data movement by $5$ orders of magnitude, while dividing by \maximed{$2.3$} the 
time required to analyze a time-step (line 
\emph{``Analysis time''}, \tabref{tab:performance}).}




\subsection{\maxime{Ranking quality}}

\maxime{In this section, we evaluate \julien{quantitatively} the 
\julien{relevance} of \julien{the} rankings obtained
with each of the metrics discussed in \secref{sec:metrics},
and compare them to rankings obtained with
\maximed{\emph{overlap methods}, traditionally used for associating
geometrical sub-domains \cite{Saikia17, bremer10, bremer_tvcg11, bajaj06, 
Silver95}.}
}
\maximed{Let \julien{$\mathcal{F}_{A_t}$ be the acquired finger 
geometry (\secref{sub:protocol}) 
and $\mathcal{F}_{S_t}$ be the sub-level set of the simulated water saturation 
at time  $t$.}
The overlap $O(A_t,S_t)$ between $A_t$ and $S_t$ is the volume
of $\mathcal{F}_{A_t} \cap \mathcal{F}_{S_t}$
divided by the volume of $\mathcal{F}_{A_t} \cup \mathcal{F}_{S_t}$.
From this we can define a distance:}

\vspace{-2ex}
\begin{itemize}
  \item{$d_O(A,S) = \Big(\sum_t \big(1 - O(A_t,S_t)\big)^2 \Big)^{1/2}$}
\end{itemize}
\vspace{-2ex}

\noindent
\maximed{Integrating the overlap $\otherOverlap{t}(S)=1-O(S_t,S_{t+1})$ between 
$S_t$ and $S_{t+1}$
for a single simulation and comparing it to the integrated overlap for the acquisition
yields a \emph{velocity}-oriented version:}

\vspace{-2ex}
\maximet{
\begin{itemize}
  \item{$d_{\otherOverlapIndex{}}(A,S) = 
   \Big(\sum_t \big(\otherOverlap{t}(S) - \otherOverlap{t}(A)\big)^2 \Big)^{1/2} $}
\end{itemize}}
\vspace{-1ex}

\maxime{\julien{At this point, w}e \maximed{need to} compare 
different rankings to the reference ranking constituted by experts.
Let $R_1$ and $R_2$ be two rankings of $n$ simulations.
One of the most commonly used methods for computing a degree of
similarity between $R_1$ and $R_2$ is \textbf{Kendall's $\tau$}
\cite{croux2010influence, goktas2011comparison}:
for all couples $(\maximej{r}_i,\maximej{r}_j) \in R_1^2$ and 
$(\maximej{s}_i,\maximej{s}_j) \in R_2^2$:}

\vspace{-3ex}
\begin{eqnarray}
 \tau = \frac{2}{n(n-1)} \sum_{i<j} \text{sign}(\maximej{r}_i-\maximej{r}_j) \text{sign}(\maximej{s}_i-\maximej{s}_j)
\end{eqnarray}
\vspace{-3ex}

\noindent
\maxime{It corresponds to the number of pairs $(i,j)$
for which $\maximej{r}_i$ and $\maximej{r}_j$ in $R_1$ have the same ordering as
$\maximej{s}_i$ and $\maximej{s}_j$ in $R_2$ minus the number of pairs
for which the orderings in $R_1$ and $R_2$ are different.
In other words, it is the difference between the number
of concordant pairs and the number of discordant pairs.
The closer this number is to $1$ in absolute value,
the more compatible the rankings\maximed{, 
\maximej{$\tau$ being close to $-1$ indicates} that the two rankings are
in reverse order.}}

\maxime{Over the 200 examined runs,
many were quickly discarded by experts during the manual ranking,
because they were too far from the acquisition. Thus,
as the order of the poorest runs is not important to the experts,
we also compute the similarity with the reference ranking for the best 
(top 50 and top 25) identified runs according to each method.
\julienLDAV{This enables us to focus the evaluation of the performance of our 
framework in separating plausible from non-plausible runs.}
Resulting Kendall coefficients are exposed in \tabref{tab:quality}.}

\begin{table}
    \centering
    \caption{
      Quality of rankings. Kendall coefficients
      between each ranking and the reference ground truth
      formed by experts are computed \maximet{(closest to $1$ is best)}. 
      Since the order in which
      the poorest runs are ordered in the expert's ranking is
      arbitrary, coefficients are also computed for the (50 and 25) 
      best simulations according to each method. 
      \maximet{The best coefficient for each case is shown in bold.}
    }
    \label{tab:quality}
    \scalebox{0.85}{
    \centering
    \begin{tabular}{|l|rrrr|rr|}
    \hline 
    \rule{0pt}{1.0\normalbaselineskip}Method
    & \multicolumn{1}{c|}{$O$}
    & \multicolumn{1}{c|}{$\wasserstein{2}$}
    & \multicolumn{1}{c|}{$\liftedWasserstein{2}$}
    & \multicolumn{1}{c|}{$EMD$}
    & \multicolumn{1}{c|}{$\otherOverlap{}$}
    & \multicolumn{1}{c|}{$\otherLiftedWasserstein{2}$} \\
    \hline 
    \rule{0pt}{1.0\normalbaselineskip}All
             & 0.37 & 0.25 & 0.26          & 0.15 & 0.12  & \textbf{0.41} \\
    Top 50   & 0.22 & 0.46 & \textbf{0.66} & 0.47 & -0.29 & 0.46 \\
    Top 25   & 0.13 & 0.29 & \textbf{0.84} & 0.70 & -0.13 & 0.42 \\
    \hline
    \end{tabular}
    }
    \vspace{-2ex}
\end{table}

\maximet{Observing lines 2 and 3 in \tabref{tab:quality}, 
we can first note that the overlap method (column ``$O$'') does not 
perform well. This behavior was expected because of the very
chaotic geometry of fingers.
The Wasserstein method, which is the traditional reference
metric for comparing persistence diagrams, is shown in column ``$W_2$''. 
The Earth mover's distance method (column ``$EMD$'') only takes the 
geometrical information of extrema into account, regardless of their
persistence. It seems to perform better than $W_2$, which is 
unexpected, because $EMD$ can wrongly associate small-scale details to
large-scale ones.
The lifted Wasserstein method, which 
includes persistence information and favors a geometrical direction, 
is shown in column ``$\liftedWasserstein{2}$''.
As it achieves the best overall Kendall coefficients, it seems that 
$\liftedWasserstein{2}$ manages to combine the advantages of both 
$EMD$ and $W_2$, not just being a simple \julienLDAV{interpolation} between the 
two.
Lastly, metrics based on the 
distances traveled by fingers
(columns $\otherOverlap{}$ and $\otherLiftedWasserstein{2}$) 
do not appear to be able to produce relevant rankings.}


\begin{table}[h]
    \centering
    \caption{
      \julienLDAV{Appreciation of the top-25 rankings returned by each method.}
      Diffuse runs, slow runs and runs in common with
      the expert's ranking are counted for each method.
      Each ranking is shown to an expert using our web
      interface and their appreciation is noted.
    }
    \label{tab:appreciation}
    \scalebox{0.85}{
    \centering
    \begin{tabular}{|l|rrrr|rr|}
    \hline 
    \rule{0pt}{1.0\normalbaselineskip}Method
    & \multicolumn{1}{c|}{$O$}
    & \multicolumn{1}{c|}{$\wasserstein{2}$} 
    & \multicolumn{1}{c|}{$\liftedWasserstein{2}$} 
    & \multicolumn{1}{c|}{$EMD$} 
    & \multicolumn{1}{c|}{$\otherOverlap{}$} 
    & \multicolumn{1}{c|}{$\otherLiftedWasserstein{2}$} \\
    \hline 
    \rule{0pt}{1.0\normalbaselineskip}too diffuse
                  & 0    & 4    & 0    & 5    & 0     & 7     \\
    too slow      & 0    & 0    & 0    & 0    & 17    & 0     \\
    common        & 0    & 10   & 21   & 18   & 0     & 11    \\
    appreciation  & poor & good & best & poor & wrong & wrong \\
    \hline
    \end{tabular}
    }
\end{table}




\subsection{\julien{Expert feedback}}

\begin{figure}
 \centering
 \includegraphics[width=\columnwidth]{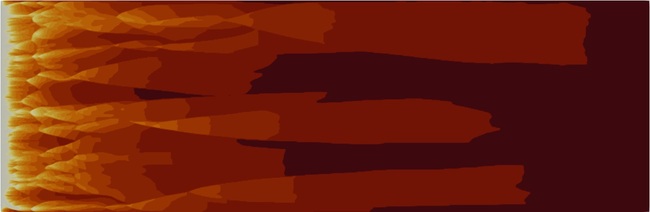}
 \vcaption{\maximet{Diffuse run example. The tips of fingers 
 grow wider than their base, forming a sort of inverted funnels. 
 The saturation field does not exhibit a very sharp frontier with 
 the background.} } 
 \vspace{-2ex}
 \label{fig:fig14}
\end{figure}

\maximed{In this section we expose a qualitative appreciation, 
collected from experts, and a discussion of ranking results. }
\maximed{We show in \tabref{tab:appreciation} the
qualitative appreciation of rankings.
The poor performance of velocity-based
metrics ($\otherOverlap{}$ and $\otherLiftedWasserstein{2}$) was unexpected.
Looking at the rankings, we see that aberrant runs are considered close
to the ground truth by these two metrics. There are three types of aberrant
runs: too slow, too fast, and too diffusive (i.e. whose finger tips grow
large and do not form a very sharp frontier with the background, as
illustrated in \figref{fig:fig14}). 
$\otherOverlap{}$ gives a good score to runs that are too slow, and
$\otherLiftedWasserstein{2}$, on the contrary, scores highly runs
that are too diffusive.}

\maximed{The number of slow runs is counted in \tabref{tab:appreciation}.
The $\otherOverlap{}$ approach, 
\julienLDAV{based on}
overlaps in consecutive
time-steps, does not discard them. The reason for this is that in simulations,
the water saturation front is very smooth
though in the acquisition fingers display a quite dendritic structure.
\julienLDAV{Thus}, the overlap between successive time-steps of smooth fingers 
going slow 
compares close to the overlap between successive time-steps of thin fingers going fast.}
\maximed{The number of runs which exhibit a very diffusive behavior is also counted.
These diffuse fingers seem
to give trouble to the $\otherLiftedWasserstein{2}$ metric 
(and also to $EMD$ and $\wasserstein{2}$).
This is because in the set of available simulations, among all which are diffuse some
inevitably end up at the exact same advancement as the acquisition
when the threshold stage (\secref{sec:features}) is applied.
Taking into account the number of fingers ($\otherLiftedWasserstein{2}$)
or considering their branching events ($\wasserstein{2}$) is apparently insufficient
to discard them.}
\maximed{Note that the $\liftedWasserstein{2}$ metric (and even $\wasserstein{2}$)
were well appreciated by the experts because the top simulations in their rankings
display fingers whose tips are quite close to the acquisition near breakthrough time,
though for $\wasserstein{2}$, there seems to be a higher distance variability. 
As for the basic overlap method $O$, it fails to identify the real best simulations,
though it does not incorrectly bring out aberrant runs either (be it too diffuse
or too slow). Its ranking, though, feels random to the experts.
Overall, the best performing metric seems to be $\liftedWasserstein{2}$, as confirmed
quantitatively and qualitatively.}

\begin{figure}
 \centering
 \includegraphics[width=\columnwidth]{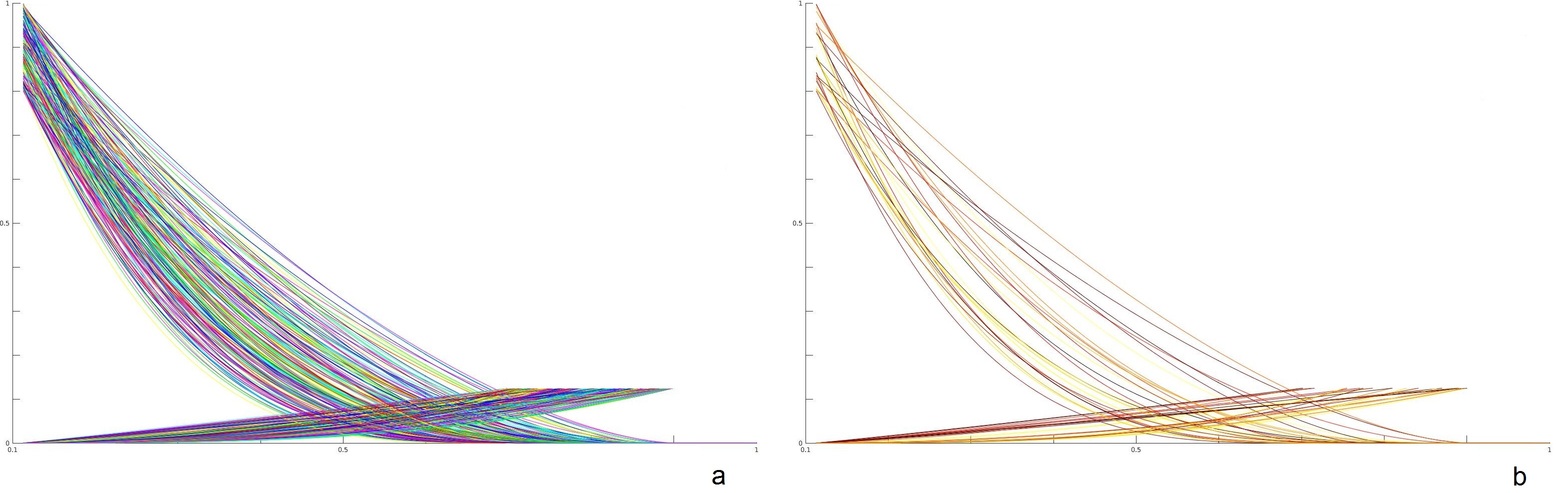}
 \vcaption{\maximet{Input relative permeability curves of all 200 simulations 
(left,
 with random colors) and for the 25 best selected runs (right, darker is \maximej{closest to ground truth}). 
 No clear pattern was seen that could discriminate relative permeability curves
 yielding the best fingers.} }
 \vspace{-2ex} 
 \label{fig:fig13}
\end{figure}

\maximed{Taking a step back, the approach we proposed is appealing to experts 
because it allows them to include geometrical information into
their parametric studies, in an autonomous and systematic
way (instead of manually inspecting and checking runs).}
\maximed{Using the $\liftedWasserstein{2}$-based ranking results,
we present in \figref{fig:fig13} all permeability curves and those yielding the
best simulation runs. We see no clear pattern arising, either visually
or numerically with respect to $L_2$ or Haussdorff \maximej{\cite{munkres2014topology}} 
distances between curves. 
This \maximej{confirms} the well-known difficulty of calibrating $kr$ curves.}

\section{Conclusion}

\maximet{In this \julienLDAV{application} paper, we presented a framework 
for enabling the automatic comparison and ranking of 
simulation runs
to an acquired ground truth. We presented a set of metrics 
specifically adapted to this 
task in the case of viscous fingering in porous media. 
After evaluation, we identified the
best fitting approach (the $\liftedWasserstein{2}$ metric, which computes
a geometrically tuned Wasserstein distance between simulation
and acquisition \julienWed{persistence diagrams}, on a per-time-step basis).
This quantitative measurement method supplements the expert's, and 
allows them to automatically form a subjective ranking close to one 
they would have manually produced. 
We demonstrated the possibility and showed the advantage of
implementing the computation of this metric in-situ\maximej{, speeding up
the analysis pipeline by a factor of 2.3 and reducing
data movement by $5$ orders of magnitude}.
We proposed a lightweight web interface to explore automatically 
generated rankings and manually edit them.}
\maxime{As with \maximed{the best metric} $\liftedWasserstein{2}$, 
\maximed{there are still some diffuse runs in the ranked best fifty, }
we believe \maximed{it} 
could be further enhanced, for instance by considering the sharpness
of the water saturation front, or by augmented the persistence diagram
with the individual volume of fingers.} 
\maximej{Besides, though in our experimental setting, 
simulations took place in a 2D domain, nothing in 
our approach is restrictive to this case. 
\julienLDAV{In future work, our framework could be experimented with simulation 
models other than Darcy's. Moreover, it could also be applied to 3D cases.}
\julienLDAV{Our overall approach would be usable as-is in these scenarios, but 
its meta-parameters (water saturation threshold and lifting coefficients) would 
likely need to be adjusted. 
\maximeLDAV{Automatically optimizing the values of these 
meta-parameters is also a promising direction for future work.} Furthermore, the 
metrics introduced in this work have been mostly motivated empirically based on 
interactions with domain experts. In the future, we will consider a theoretical 
investigation of the stability of these metrics (based on stability results on 
the Bottleneck \cite{cohen-steiner05} and Wasserstein \cite{Cohen-SteinerEHM10} 
metrics).}}
\maximeLDAV{Further, ways to capture finger merging events, 
for instance by considering richer topological structures such as
Reeb Graphs, could also be considered, as this seems to happen in 
acquired images.} 

\maxime{On another note, on the set of 200 simulations, 
we were not able to identify a 
regime of best-matching input parameters.
The combination of our metric with production and
pressure data (at injectors/producers) in a follow-up study
would be interesting in this regard. 
Trying to understand the influence of the space of
\maximet{input} parameters, here $kr$ curves, proves quite challenging. 
In our study, only four parameters (power law exponents and endpoints)
were sampled, yielding a four-dimensional space, 
but the number of sampled parameter may be significantly higher.
\maximej{In particular, this study may be extended to 
permeability curves other than based on simple Corey power laws.} 
We think it would be insightful to
develop a visual interface for exploring \maximet{such spaces of 
model parameters. We hope to see, in future studies, 
how accounting for the geometrical and topological quality of a
modeled phenomenon can be used to infer or restrict model parameters.}}




%

\acknowledgments{
\small{
\maximeLDAV{
This work is partially supported by the Bpifrance grant “AVIDO” (Programme 
d’Investissements d’Avenir, reference P112017-2661376/DOS0021427), by the 
French National Association for Research and Technology (ANRT), in the framework
of the LIP6 - Total SA CIFRE partnership reference 2016/0010 and by the 
European Commission grant H2020-FETHPC “VESTEC” (ref. 800904). 
The authors would like
to thank the anonymous reviewers for their thoughtful remarks and suggestions.}
}
}

\clearpage

\bibliographystyle{abbrv-doi}

\bibliography{paper}
\end{document}